\documentstyle[epsfig,aps,prb]{revtex}
\newcommand\beq{\begin{equation}}
\newcommand\eeq{\end{equation}}
\newcommand\bea{\begin{eqnarray}}
\newcommand\eea{\end{eqnarray}}
\newcommand\non{\nonumber}
\newcommand\bib{\bibitem}

\begin{document}

\begin{center}
{\Large Study of an Antiferromagnetic Sawtooth Chain \\ with Spin-1/2 and 
Spin-1 sites}
\end{center}

\vskip .5 true cm
\centerline{\bf V. Ravi Chandra and Diptiman Sen}
\vskip .5 true cm

\centerline{\it Centre for Theoretical Studies, Indian Institute of 
Science,}
\centerline{\it Bangalore 560012, India}
\vskip .5 true cm

\centerline{\bf N. B. Ivanov \cite{paddress} and J. Richter}
\vskip .5 true cm

\centerline{\it Institut f\"ur Theoretische Physik, Universit\"at 
Magdeburg,}
\centerline{\it P.O. Box 4120, D-39016 Magdeburg, Germany}
\vskip .5 true cm

\begin{abstract}
We study the low-energy properties of a sawtooth chain with spin-1's at the 
bases of the triangles and spin-1/2's at the vertices of the triangles. The 
spins have Heisenberg antiferromagnetic interactions between nearest neighbors,
with a coupling $J_2$ between a spin-1 and a spin-1/2, and a coupling $J_1 =1$ 
between two spin-1's. Analysis of the exact diagonalization data for periodic 
chains containing up to $N=12$ unit cells shows that the ground state is a 
singlet for exchange couplings up to approximately $J_2=3.8$, whereas for 
larger $J_2$, the system exhibits a ferrimagnetic ground state characterized 
by a net ferromagnetic moment per unit cell of 1/2. In the region of small 
interactions $J_2$, the mixed spin sawtooth chain maps on to an effective 
isotropic spin model representing two weakly interacting and frustrated
spin-1/2 Heisenberg chains composed of spin-1/2 sites at odd and even vertices
respectively. Finally, we study the phenomenon of a macroscopic magnetization 
jump which occurs if a magnetic field is applied with a value close to the 
saturation field for $J_2 =2$.

\end{abstract}
\vskip .5 true cm

~~~~~~ PACS number: ~75.10.Jm, ~75.50.Ee, ~75.30.Ds, ~75.45.+j

\newpage

\section{Introduction}

There has been a great deal of interest in recent years in one-dimensional 
quantum spin systems with frustration \cite{lhuillier}. The most common
examples of such systems are those in which triangles of Heisenberg spins 
interact antiferromagnetically with each other. Some of the systems which
have been studied analytically or numerically so far are the sawtooth spin-1/2
chain \cite{nakamura,blundell}, a chain of spin-1/2 triangles \cite{raghu}, 
frustrated mixed spin ferrimagnetic chains \cite{ivanov}, and the spin-1/2 
Kagome strip \cite{azaria,waldtmann}. 
 
Classically (i.e., in the limit in which the magnitudes of the spins $S_i 
\rightarrow \infty$), some of these frustrated systems have an enormous 
ground state degeneracy arising from local rotational degrees of freedom 
which cost no energy. Quantum mechanically, this degeneracy is often lifted 
due to tunneling between different classical ground 
states. However, one might still expect a remnant of the classical degeneracy 
in the form of a large number of low-energy excitations in the quantum system. 

Recently, the ground state of the spin-1/2 sawtooth chain has been numerically
studied as a function of the ratio $J_2/J_1$, where $J_1$
is the coupling between pairs of spins at the bases of the triangles, and $J_2$
is the coupling between a spin at the base and a spin at the vertex of a 
triangle \cite{blundell}. The system was found to be gapless for $J_2/J_1 > 
2.052$ and for $J_2/J_1 < 0.65$. The low-energy excitations have the
same dispersion for singlets and triplets. For $J_2/J_1 =1$, the system has
some special properties. The ground state of an open chain has an exact 
degeneracy which increases linearly with the number of triangles 
\cite{nakamura}. This degeneracy arises from the existence of localized
spin-1/2 kinks which do not cost any energy regardless of their position in 
the chain. There are also spin-1/2 antikinks which cost a finite energy. The 
lowest excitation in a chain with periodic boundary conditions is 
given by a kink-antikink pair which has a dispersionless gap; the pair may 
be either a singlet or triplet.

In this paper, we will carry out analytical and numerical studies of a mixed 
spin Heisenberg antiferromagnet on the sawtooth lattice shown in Fig. 1. (The 
arrows and angles ($\theta$) shown in that figure refer to a canted state which
will be discussed later). The sites at the vertices of the triangles have spin 
$S_2$, and they are labeled $1,2,...,N$. The sites at the bases of the 
triangles have spin $S_1$, and they are labeled $N+1,N+2,...,2N$. The 
number of triangles is therefore $N$. The Hamiltonian governing the system is
\beq
H ~=~J_1 \sum_{i=N+1}^{2N} ~{\vec S}_i \cdot {\vec S}_{i+1} ~+~ 
J_2 \sum_{i=1}^N ~{\vec S}_i \cdot [~{\vec S}_{i+N} + {\vec S}_{i+N+1})~]~,
\label{ham}
\eeq
where the couplings $J_1, J_2$ are positive. It is convenient to set $J_1 =1$,
and consider the properties of the system as a function of $J_2$. 
We will impose periodic boundary conditions at the ends of the chain, so that 
the momentum is a good quantum number. (We will set Planck's constant $\hbar 
=1$, and the nearest neighbor lattice spacings equal to 1).

The plan of our paper is as follows. In Sec. II, we will develop the spin 
wave theory (SWT) for this system \cite{anderson}, taking the values of the 
spin $S_1$ at the bases and the spin $S_2$ at the vertices of the triangles 
to be very large, and $S_1 > S_2$. If $J_2 > 2S_1/S_2$, we find that the 
system is a ferrimagnet, with a magnetization per unit cell of $S_1 - S_2$. 
If $J_2 < 2S_1/S_2$, we find that there is an infinite number of classical 
ground states as mentioned above. For reasons explained below, we will 
consider the classical ground states which are 
coplanar; the number of even this restricted set of states grows exponentially
with $N$. We perform a linear SWT about these coplanar states, and find that 
the spin wave zero point energy does not break the classical degeneracy. 
Further, one of the spin wave modes turns out to have zero energy for all
momenta. We will also see that SWT picks out two other values of $J_2$, 
namely $J_2 =1$ and 2, as being special.

In Sec. III, we use the Lanczos algorithm to perform an exact diagonalization 
(ED) of finite systems to study the low-energy excitations 
and two-spin correlations in the ground state as a function of $J_2$ for
$S_1 =1$ and $S_2 =1/2$. We find that the system is a ferrimagnet for $J_2
\gtrsim 3.8$ with a magnetization per unit cell of 1/2. [We see that the 
transition to a collinear ferrimagnetic state takes place at a smaller value 
of $J_2$ in the quantum case than in the classical case where the transition
occurs at $J_2 =4$. This effect has already been seen for other systems 
exhibiting transitions between collinear and non-collinear states (see, for
example, Ref. \cite{ivanov}), and it indicates a favoring of the collinear 
state by quantum fluctuations]. There seems to be 
a first-order transition at $J_2 \simeq 3.8$ with the total spin of the ground
state changing rather abruptly at that value. For $J_2 \lesssim 3.8$, the 
ground state is a singlet. We find that there are two other values, $J_2 
\simeq 1.9$ and 1.1, where the nature of the spin correlations changes 
significantly. Many of the correlations become very small or change sign at 
those two points. The structure factor seems to indicate crossovers at those 
points between ground states with different kinds of short-range correlations.
In the region $1.1 \lesssim J_2 \lesssim 1.9$, the canted spin
configuration in Fig. 1 is consistent with the ED data representing the
short-range spin-spin correlations, whereas for larger $J_2$ up to
the ferrimagnetic phase transition point, the commensurate spiral phase
with a period of four lattice spacings seems to be in accord to the
ED data for $N=12$. It is clear, however, that the periodic boundary 
conditions imposed on the chain prevent the appearance of the periodic 
structures with larger periods predicted by the classical analysis.

For $J_2 \lesssim 1$,
the correlations between the spin-1/2 sites show an unusual pattern. Namely, 
the spin-1/2 sites appear to decompose into two sublattices such that each 
sublattice has a substantial antiferromagnetic coupling within itself
(with a strong frustration), but the coupling between the two sublattices is 
much weaker. We call this system the next-nearest-neighbor antiferromagnet 
(NNN-AFM). In Sec. IV, we use a perturbative expansion in $J_2$ and an 
effective Hamiltonian description to provide some understanding of why this 
happens. This seems to be a remarkable property of the spin-1/2 - spin-1
sawtooth system.

In Sec. V, we will consider the particular case of $J_2 = 2$ where we find
that the system shows an interesting behavior if a magnetic field is applied
with a strength which is close to the saturation value $h_s$ (i.e., the value 
above which all the spins are aligned with the field). We will show that for 
$J_2 = 2$, the system displays a macroscopic jump in the magnetization as the 
magnetic field crosses $h_s$. This phenomenon is known to occur in some other 
strongly frustrated quantum spin systems \cite{schulenburg,schnack}.

\section{Spin wave analysis}

To develop the SWT, we assume that the values of the spin $S_1$
and $S_2$ are much larger than 1. We will describe how to obtain the spin wave
dispersion up to order $S_i$. (This is called linear SWT because interactions 
between the spin waves do not appear at this order).

Since some of the classical ground states considered in this section have
a coplanar configuration of the spins, it is convenient to use a technique for
deriving the spin wave spectrum which can be applied to both collinear and
coplanar configurations. For a coplanar configuration, let us assume that
the spins lie in the $z-x$ plane. Consider a particular spin of magnitude
$S$ which points at an angle $\phi$ with respect to the $\hat z$-direction.
Then we can write the Holstein-Primakoff representation for that spin as
\bea
\cos \phi ~S_z ~+~ \sin \phi ~S_x &=& S ~-~ a^\dagger a ~, \non \\
- ~\sin \phi ~S_z ~+~ \cos \phi ~S_x ~+~ i S_y &=& {\sqrt {2S ~-~ 
a^\dagger a}} ~a ~, \non \\
- ~\sin \phi ~S_z ~+~ \cos \phi ~S_x ~-~ i S_y &=& a^\dagger ~{\sqrt {2S ~
-~ a^\dagger a}} ~,
\label{hp1}
\eea
where $[a,a^\dagger ] =1$. We now introduce a coordinate and momentum
\beq
q ~=~ \frac{a ~+~ a^\dagger}{\sqrt 2} ~, \quad {\rm and} \quad 
p ~=~ \frac{a ~-~ a^\dagger}{i {\sqrt 2}} ~,
\eeq
satisfying $[q,p]=i$. On expanding Eq. (\ref{hp1}) up to quadratic order 
in $a$ and $a^\dagger$, we obtain
\bea
S_z &=& \cos \phi ~[~ S ~+~ \frac{1}{2} ~-~ \frac{1}{2} ~(p^2 + q^2) ~]~ -~
\sin \phi ~{\sqrt S} ~q ~, \non \\
S_x &=& \sin \phi ~[~ S ~+~ \frac{1}{2} ~-~ \frac{1}{2} ~(p^2 + q^2) ~]~ +~
\cos \phi ~{\sqrt S} ~q ~, \non \\
S_y &=& {\sqrt S} ~p ~.
\label{hp2}
\eea

We now consider a general Heisenberg Hamiltonian of the form 
\beq
H ~=~ \sum_{ij} ~J_{ij} ~{\vec S}_i \cdot {\vec S}_j ~,
\eeq
where we count each bond $(ij)$ only once, and the spin at site $i$ will
be assumed to have a magnitude $S_i$. Consider a classical 
configuration in which the spin at site $i$ lies in the $z-x$ plane
at an angle $\phi_i$ with respect to the $\hat z$-axis. The condition for this
configuration to be a ground state classically is that 
\beq
E_{cl} (\phi_i) ~=~ \sum_{ij} ~J_{ij} ~S_i S_j ~\cos (\phi_i - \phi_j)
\label{ecl}
\eeq
should be a minimum with respect to each of the angles $\phi_i$. We must
therefore have
\beq
\sum_j ~J_{ij} ~S_i S_j ~\sin (\phi_i - \phi_j) ~=~ 0 
\eeq
for every value of $i$. Using Eq. (\ref{hp2}) and keeping terms up to order
$S_i$, we find that the spin wave Hamiltonian is given by 
\bea
H_{sw} ~=~ \sum_{ij} ~J_{ij} ~[& & (S_i S_j + \frac{S_i}{2} + \frac{S_j}{2}) ~
\cos (\phi_i - \phi_j) ~-~ \frac{1}{2} ~\cos (\phi_i - \phi_j) ~(S_j p_i^2 + 
S_j q_i^2 + S_i p_j^2 + S_i q_j^2) \non \\
& & +~ {\sqrt {S_i S_j}} ~ \cos (\phi_i - \phi_j) ~q_i q_j ~+~ {\sqrt {S_i 
S_j}} ~p_i p_j ~] ~.
\label{hamsw1}
\eea
The factor of $S_i S_j + S_i/2 + S_j/2$ in this expression appears on
expanding a product like $(S_i +1/2)(S_j +1/2)$ coming from Eq. (4) and
dropping the term of order 1.

We can obtain the spin wave spectrum from (\ref{hamsw1}) as follows. The
unit cell of our system is a triangle containing the two sites with spins 
$S_1$ and $S_2$ which lie on its left edge. Let us label the triangles by 
$n$, where $n=1,2,...,N$, and let $a=1,2$ denotes the spins $S_1$ and $S_2$ 
respectively; thus each site is labeled as $(a,n)$. [The mapping from the 
site labels used in Fig. 1 to the site labels $(a,n)$ being used here is as 
follows: $n \rightarrow (2,n)$ if $1 \le n \le N$, and $n \rightarrow (1,n-N)$
if $N+1 \le n \le 2N$]. We define the Fourier transforms
\bea
p_{a,k} &=& \frac{1}{\sqrt N} ~\sum_n ~p_{a,n} ~ e^{-ikn} ~, \non \\ 
q_{a,k} &=& \frac{1}{\sqrt N} ~\sum_n ~q_{a,n} ~e^{-ikn} ~, 
\eea
where $-\pi < k \le \pi$. These operators satisfy the commutation relation 
$[q_{a,k}, p_{b,k^\prime} ] 
= i \delta_{ab} \delta_{k,-k^\prime}$. Let us now assume that the cosines 
appearing in Eq. (\ref{hamsw1}) take the following simple forms: they are 
equal to $\cos \alpha$ for every pair of neighboring spin-$S_1$ sites, and 
equal to $\cos \beta$ for every pair of neighboring spin-$S_1$ - spin-$S_2$ 
sites. [We will see below that this may happen even in situations where the 
angles $\phi_{a,n}$ are themselves not the same in all the triangles]. 
Up to terms of order $S_i$, the Hamiltonian in Eq. (\ref{ham}) takes 
the form
\bea
H &=& E_{0,cl} ~+~ \sum_{ab} \sum_{\vec k} ~[~ p_{a,-k} M_{ab,k}
p_{b,k} ~+~ q_{a,-k} N_{ab,k} q_{b,k} ~]~, \non \\
E_{0,cl} &=& N ~[~ (S_1^2 + S_1) ~\cos \alpha ~+~ 2 J_2 ~(S_1 S_2 +
\frac{S_1}{2} + \frac{S_2}{2}) ~\cos \beta ~]~,
\label{hamsw2}
\eea
where $E_{0,cl}$ is the classical ground state energy, and
\bea
M_{ab,k} &=& \left( \begin{array}{cc} S_1 \cos k - S_1 \cos \alpha
- J_2 S_2 \cos \beta & ~~~~ J_2 {\sqrt {S_1 S_2}} (1 + e^{-ik})/2 ~~~~ \\
& \\
J_2 {\sqrt {S_1 S_2}} (1 + e^{ik})/2 & - J_2 S_1 \cos \beta \end{array}
\right) ~, \non \\
& & \non \\
N_{ab,k} &=& \left( \begin{array}{cc} S_1 \cos \alpha \cos k - S_1 
\cos \alpha - J_2 S_2 \cos \beta & ~~~~ J_2 {\sqrt {S_1 S_2}} \cos \beta (1 + 
e^{-ik})/2 ~~~~ \\
& \\
J_2 {\sqrt {S_1 S_2}} \cos \beta (1 + e^{ik})/2 & - J_2 S_1 \cos \beta
\end{array} \right) ~.
\label{mn}
\eea
Note that the $2 \times 2$ matrices ${\underline M}_k$ and ${\underline N}_k$ 
satisfy ${\underline M}_{-k} = {\underline M}_k^T$ and ${\underline N}_{-k} 
= {\underline N}_k^T$. If we write $p_{a,k}$ and $q_{a,k}$ as the columns 
${\underline p}_k$ and ${\underline q}_k$ respectively, then the classical
Hamiltonian equations of motion take the form
\bea
\frac{d{\underline q}_k}{dt} &=& 2 {\underline M}_k {\underline p}_k ~, 
\non \\
\frac{d{\underline p}_k}{dt} &=& - 2 {\underline N}_k {\underline q}_k ~.
\eea
For each value of $k$, the harmonic solutions of these equations have two 
possible frequencies $\omega_k$ given by the eigenvalue equation
\beq
{\rm det} ~(4 {\underline M}_k {\underline N}_k ~-~ \omega_k^2 ~{\underline 
I}) ~=~ 0 ~.
\label{omega}
\eeq

The quantum mechanical energy levels are then given by $(n_{a,k} +1/2)
\omega_{a,k}$, where $n_{a,k}$ is the occupation number of the mode labeled
as $(a,k)$, where $a$ can take two different values. Note that the frequencies
$\omega_{a,k}$ are the same in all the coplanar configurations. Hence the
zero point energy given by $(1/2) \sum_{a,k} \omega_{a,k}$ does not
break the classical degeneracy between the different configurations.

[The advantage of using the variables $p$ and $q$, instead of $a$ and 
$a^\dagger$ is the following. We never get cross-terms like $p_i q_j$ in the 
Hamiltonian; hence it is straightforward to obtain the frequencies. On 
the other hand, if we use $a$ and $a^\dagger$, we get terms like $a_i a_j$ 
which make it necessary to perform a Bogoliubov transformation to obtain the
frequencies. It turns out, however, that in both cases we eventually have to 
diagonalize the same matrix (namely, ${\underline D}_k = 4 {\underline M}_k 
{\underline N}_k$) to obtain the frequencies]. 

We can now obtain the spin wave dispersion for various values of $J_2$.
For large values of $J_2$, the classical ground state is a collinear 
ferrimagnetic configuration in which the $S_1$ spins point in one direction, 
say, the $\hat z$-direction, and the $S_2$ spins point in the opposite 
direction; the total spin of the ground state is therefore equal to $N(S_1 - 
S_2)$. Hence the cosines in Eq. (\ref{mn}) are given by $\cos \alpha =1$ and 
$\cos \beta = -1$. The spin wave dispersions are then given by
\bea
\omega_{\pm,k} &=& 2 {\sqrt {a_k^2 - c_k^2}} ~\pm ~2 b_k ~, \non \\
a_k &=& \frac{J_2}{2} ~(S_1 + S_2) ~-~ S_1 \sin^2 \left( \frac{k}{2} 
\right) ~, \non \\
b_k &=& \frac{J_2}{2} ~(S_1 - S_2) ~+~ S_1 \sin^2 \left( \frac{k}{2}
\right) ~, \non \\
c_k &=& J_2 {\sqrt {S_1 S_2}} \cos \left( \frac{k}{2} \right) ~.
\label{swf}
\eea
[We can show that the upper branch $\omega_{+,k}$ corresponds to 
excitations with total spin one more than the ground state spin, while
the lower branch $\omega_{-,k}$ corresponds to excitations with total spin 
one less than the ground state spin]. These dispersions are shown in Fig. 2 
for $J_2 =5$, $S_1 =1$ and $S_2 =0.5$.
At $k=0$, we find that $\omega_{+,0} = 2J_2 (S_1 - S_2)$ and $\omega_{-,0} =0$.
At $k=\pi$, $\omega_{+,\pi} = 2J_2 S_1$ while $\omega_{-,\pi} =
2J_2 S_2 - 4 S_1$. When the ratio $J_2$ decreases to the value $2S_1/S_2$,
the lower branch $\omega_{-,k}$ vanishes for all values of $k$. This signals
an instability to some other state for $J_2 < 2S_1 /S_2$.

For later use, we note that up to order $S_i$, the ground state energy per 
unit cell in the ferrimagnetic phase is given by
\bea
\frac{E_0}{N} = & & E_{0,cl} ~+~ \frac{1}{2} ~\int_{-\pi}^\pi ~
\frac{dk}{2\pi} ~[~ \omega_{+,k} ~+~ \omega_{-,k} ~] \non \\
= & & S_1^2 ~+~ S_1 ~-~ 2J_2 ~(S_1 S_2 + \frac{S_1}{2} + \frac{S_2}{2}) ~+~ 
\int_0^\pi ~\frac{dk}{2\pi} ~4 ~{\sqrt {a_k^2 ~-~ c_k^2}} ~, 
\label{e0f}
\eea
where $a_k$ and $c_k$ are given in Eq. (\ref{swf}).

For $J_2 < 2 S_1 /S_2$, the classical ground state is no longer a collinear
state. To see this, note that the Hamiltonian in (\ref{ham}) can be written,
up to a constant, as $H = (1/2) \sum_n {\vec W}_n^2$, where
\beq
{\vec W}_n ~=~ J_2 ~{\vec S}_{2,n} ~+~ {\vec S}_{1,n} + {\vec S}_{1,n+1} ~.
\label{wn}
\eeq
Thus the classical 
ground state is one in which the vector ${\vec W}_n$ has the minimum
possible magnitude in each triangle $n$. For $J_2 < 2S_1/S_2$, we find that
the lowest energy state in each triangle is one in which the magnitude of 
${\vec W}_n$ is zero; this is given by a configuration in which the spin-$S_2$
makes an angle of $\pi - \theta$ with both the spin-$S_1$'s, while the angle 
between the two spin-$S_1$'s is $2\theta$, where 
\beq
\theta ~=~ \cos^{-1} \left( \frac{J_2S_2}{2S_1} \right) ~.
\eeq
(Fig. 1 shows a particularly simple example of such a configuration in which
all the spin-$S_2$'s are aligned with each other; this
is called the canted state). It is 
clear that there is an infinite number of such configurations even in a system
with a finite number of triangles. This is because, in a triangle labeled $n$,
we can continuously rotate the spins $S_{2,n}$ and $S_{1,n+1}$ around the spin
$S_{1,n}$ while maintaining the relative angles at the values given above.
In many systems with such an enormous ground state degeneracy, it is known 
that the zero point energy in linear SWT breaks the degeneracy partially by 
selecting only the coplanar ground states; this is called the 
order-from-disorder phenomenon \cite{doucot}. Let us therefore consider only
coplanar configurations, in which all the spins lie in the $z-x$ plane. Even 
with this restriction, there are about $2^N$ different configurations, because
in triangle $n$, there are two possible directions of the spins $S_{2,n}$ and 
$S_{1,n+1}$ for a given direction of the spin $S_{1,n}$. 

Let us compute the spin wave dispersion in a coplanar configuration. The 
cosines in Eq. (\ref{mn}) are given by
\bea
\cos \alpha &=& \cos (2 \theta) ~=~ \frac{J_2^2 S_2^2}{2S_1^2} ~-~ 1 ~, \non \\
\cos \beta &=& - ~\cos \theta ~=~ - ~\frac{J_2 S_2}{2S_1} ~.
\eea
We then find that ${\rm det} {\underline M}_k =0$ for all values of $k$. Eq. 
(\ref{omega}) then implies that one of the frequencies, say, $\omega_{-,k} = 0$
for all $k$. We thus have a dispersionless zero mode. [This mode arises due to
the invariance of the classical ground state energy under certain kinds of 
continuous rotations in each triangle as mentioned above. In the problem
of the Heisenberg antiferromagnet on the Kagome lattice, it is known that
interactions between spin waves, which appear when we go to higher orders
in the $1/S$ expansion, remove the degeneracy in the zero-mode branch 
\cite{chubukov}, and produce a low-lying spin wave branch with an energy 
scale proportional to $S^{2/3}$. We will restrict ourselves to linear SWT
here, and will not consider such corrections to the zero-mode branch].

Since $\omega_{-,k} =0$, the other frequency can be obtained from (\ref{omega})
as
\bea
\omega^2_{+,k} ~= & & 4 ~{\rm tr} ~({\underline M}_k {\underline N}_k) \non \\
= & & 2J_2^2 S_2^2 ~(\cos k ~-~ J_2) ~(1 ~+~ \cos k) ~+ ~4 S_1^2 ~\sin^2 k ~
+~ J_2^4 S_2^2 ~.
\label{swc}
\eea
This dispersion is shown in Fig. 3 for $J_2 =2$, $S_1 =1$ and
$S_2 = 0.5$. At $k =\pi$, we have $\omega_{+,\pi} = J_2^2 S_2$, while 
at $k=0$, we have $\omega_{+,0} = J_2S_2 |2-J_2|$. We thus see that the 
gap vanishes at $k=0$ if $J_2 =2$. 

Up to order $S_i$, the ground state energy per unit cell in the coplanar 
phase is given by
\beq
\frac{E_0}{N} ~=~ - ~S_1^2 ~-~ S_1 ~-~ \frac{J_2^2}{2} ~(S_2^2 + S_2) ~+~ 
\int_0^\pi ~\frac{dk}{2\pi} ~\omega_{+,k} ~, 
\label{e0c}
\eeq
where $\omega_{+,k}$ is given in Eq. (\ref{swc}). One can check that the
expressions in (\ref{e0f}) and (\ref{e0c}) match at $J_2 =2S_1 /S_2$.

Let us now comment on a special feature of the value $J_2=2$.
Within the set of $2^N$ classical coplanar ground states, the total spin of the
system can have a wide range of values depending on the exact configuration of 
the spins. We can see this by noting that the total spin can be written as 
${\vec S}_{tot} = \sum_n {\vec V}_n$, where 
\beq
{\vec V}_n \equiv {\vec S}_{2,n} + \frac{1}{2} ({\vec S}_{1,n} + 
{\vec S}_{1,n+1}) ~.
\eeq
In any of the classical ground states for $J_2 < 2S_1/S_2$, we find that the 
magnitude of this vector is given by $|{\vec V}_n| = |S_2 + S_1 \cos \beta| =
(S_2/2)|2-J_2|$. Depending on how the vectors ${\vec V}_n$ in different 
triangles add up, the total spin of the system can therefore range from 0 to 
$(NS_2/2)|2-J_2|$. However, if $J_2 =2$, we see that ${\vec V}_n$ is
proportional to ${\vec W}_n$ in (\ref{wn}); hence all the classical ground 
states have zero spin since we know that
each of the vectors ${\vec W}_n$ has zero 
magnitude. Quantum mechanically, we expect the exponentially large classical 
degeneracy to be broken by tunneling; however we would still expect an 
unusually large number of low-energy singlet excitations for $J_2 =2$.

Another special value of $J_2$ is given by $J_2 =1$. At this point, the
Hamiltonian of a single triangle is given by the square of the total spin 
${\vec S}_n = {\vec S}_{2,n} + {\vec S}_{1,n} + {\vec S}_{1,n+1}$. Thus the 
total spin of each triangle vanishes in any of the classical ground states.

\section{Numerical results}

We have used the Lanczos algorithm to study the ground state properties of 
the sawtooth chain with $S_1 =1$ and $S_2 =1/2$ for even values of $N$ from 4 
to 12 with periodic boundary conditions. 
To reduce the sizes of the Hilbert spaces, we work in subspaces with a given 
value of the total component of the spin $S_z$ and the momentum, since these 
operators commute with the Hamiltonian. If $S_z =0$, we reduce the Hilbert 
space further by working in subspaces in which the spin parity $P_s$ is equal 
to $\pm 1$; under the transformation $P_s$, the values of the spins at all the
sites are flipped from $S_{iz} \rightarrow - S_{iz}$. One can show that the
eigenvalue of $P_s$ is related to the total spin $S$ of the state by
\beq
P_s ~=~ (-1)^{N(S_1 +S_2) - S} ~.
\eeq

Fig. 4 shows the ground state energy as a function of $J_2$ for $N=8$.
The solid line shows the numerical data, while the dashed line shows the spin 
wave results (obtained from (\ref{e0c}) for $J_2 \le 4$, and from (\ref{e0f}) 
for $J_2 \ge 4$). (We do not present the data for $N=12$ since the latter are
almost indistinguishable from those presented in Fig. 4). In the inset, the 
solid lines show piecewise linear fits to the numerical data to the left and 
right of $J_2 =3.8$, while the dotted lines show the continuations of the 
same two straight lines to the right and left of $J_2 =3.8$ respectively. This
shows a small discontinuity in the slope at $J_2 \simeq 3.8$; we find that
$(1/N) ~dE_0/dJ_2 $ is equal to -1.25 and -1.45 to the left and right
respectively of $J_2 =3.8$. (These numbers agree with the nearest 
neighbor spin-1/2 - spin-1 correlations discussed in Eq. (\ref{s1s14}) and
Fig. 8 below).

For both $N=8$ and $N=12$, we find that the total spin of the ground state 
changes abruptly at $J_2 \simeq 3.8$. For $J_2 \gtrsim 3.8$, the ground state 
spin has the ferrimagnetic value of $N(S_1 -S_2)=N/2$. For $J_2 \lesssim 3.8$,
the ground state is a singlet. The number 3.8 compares reasonably with the
SWT value of $2S_1/S_2 =4$, considering that SWT is only expected to be 
accurate for large values of $S_1$ and $S_2$. The total spin of the first 
excited state however shows a more complicated behavior as $J_2$ is varied; 
this is plotted in Fig. 5 for $N=8$. For $J_2 \gtrsim 3.9$, the first excited
state has a spin of 3 as expected from the spin wave calculations. For $J_2 
\lesssim 2.9$, the first excited state is a singlet. For $2.9 \lesssim J_2 
\lesssim 3.9$, the spin of the first excited state fluctuates considerably.
The fluctuations near $J_2 \simeq 3.9$ may be due to the finite size of the
system, and they may disappear in the thermodynamic limit.

For $J_2 \le 3.8$ and $N=8$, the energy gaps between the ground state and the 
first excited state (whose spin is shown in Fig. 5) and the first non-singlet
state are plotted as functions of $J_2$ in Fig. 6; the two gaps are shown by 
stars and circles respectively. Although the gap to the first excited state
fluctuates, we see that it is particularly small near $J_2 = 1$ and $2$. These
small gaps may represent either level crossings of the ground state (as 
discussed below) or genuine low-lying singlet excitations; it is difficult to 
distinguish between these two possibilities without going to much
larger system sizes. (We note that low-lying singlet excitations are known to 
occur in the spin-1/2 Heisenberg antiferromagnet on a Kagome lattice which is 
a well known example of a highly frustrated system \cite{lecheminant}). For 
$0.5 \lesssim J_2 \lesssim 2.5$, we see that
the gap to the first excited state (which is a singlet) is typically much 
smaller than the gap to the first non-singlet state. In fact, we find that
in this range of $J_2$, there are several singlet excitations which lie below 
the first non-singlet excitation. For instance, for $N=12$, we find four 
and eight singlet excitations lying below the first non-singlet excitation at 
$J_2 =1$ and 2 respectively.

For $N=12$, the ground state has the following properties. For $J_2 \lesssim 
3.85$, the ground state is a singlet, and the parity symmetry in 
the subspace with $S_z=0$ is given by $P_s = 1$. However, the momentum $k$ of 
the ground state repeatedly changes between 0 and $\pi$. This is shown in
Table 1. We observe that there are several crossings, particularly near 
$J_2 = 1.1$ and 1.9. Repeated level crossings like this in a finite 
sized system are often a sign of a spiral phase in the thermodynamic limit
\cite{tonegawa}; we will discuss this possibility in more detail below.

\vspace{0.4cm}

\begin{center}
\begin{tabular}{|c|c|} \hline
Range of $J_2$ & ~~ Ground state ~~ \\ 
& momentum \\ \hline
$0 < J_2 < 0.95$ & $\pi$ \\
$0.95 < J_2 < 1.05$ & 0 \\
$1.05 < J_2 < 1.26$ & $\pi$ \\
$1.26 < J_2 <1.78$ & 0 \\
$1.78 < J_2 <1.82$ & $\pi$ \\
$1.82 < J_2 <1.99$ & 0 \\
$1.99 < J_2 < 3.75$ & $\pi$ \\
~~~ $3.75 < J_2 < 3.85$ ~~~ & 0 \\ \hline
\end{tabular}
\end{center}
\vspace{0.2cm}

\centerline{Table 1. The momentum of the ground state for various values of 
$J_2$, for a chain with 12 triangles.}
\vspace{0.4cm}

Next, we examine the two-spin correlations $<{\vec S}_i \cdot {\vec S}_j>$
in the ground state. These are of three types: spin-1/2 - spin-1/2, 
spin-1/2 - spin-1, and spin-1 - spin-1. These are shown in Figs. 7-9 for 
$N=12$. (We have only shown six correlations in each case. All the other
correlations are related to these by translation and reflection symmetries).
The behaviors of all the correlations show large changes near three 
particular values of $J_2$, namely, 1.1, 1.9 and 3.8. For instance, many of 
the correlations approach zero or change sign near these three values. 

It is particularly instructive to look at the nearest neighbor spin-1/2 - 
spin-1 correlation, i.e., $<{\vec S}_1 \cdot {\vec S}_{14}>$ in Fig. 8.
By the Feynman-Hellmann theorem, this is related to the derivative with
respect to $J_2$ of the ground state energy per triangle,
\beq
\frac{1}{N} ~\frac{dE_0}{dJ_2} ~=~ 2 <{\vec S}_1 \cdot {\vec S}_{14} > ~,
\label{s1s14}
\eeq
where we have used the fact that all the nearest-neighbor spin-1/2 - spin-1
correlations are equal. We can see from Fig. 8 that (\ref{s1s14})
shows a jump at $J_2 \simeq 3.8$, which indicates a first-order transition
(we know that the ground state spin changes abruptly at that point from
$N/2$ to 0 without going through any of the intermediate values). The
jump in the values of $<{\vec  S}_1 \cdot {\vec S}_{14} >$ at $J_2 \simeq
3.8$ is consistent with the jump in the slope of the ground state energy in
Fig. 4 as discussed above. At
$J_2 \simeq 1.25$ and 1.75, (\ref{s1s14}) seems to show a change of slope but 
no jump. This could indicate either a second-order transition or a crossover 
at those points; it is difficult to distinguish between these two possibilities
since a change of slope can also arise due to finite-size effects. 

For small values of $J_2$, we observe that the spin-1 - spin-1 correlations
in Fig. 9
decay rapidly with the separation $n$ between the two sites, and they also 
oscillate as $(-1)^n$. This is expected for small $J_2$ because the spin-1 
chain is only weakly coupled to the spin-1/2's; a pure spin-1 antiferromagnetic
chain exhibits a Haldane gap and a finite correlation length of about six
lattice spacings \cite{haldane,white1}. The weak coupling also explains why the
spin-1/2 - spin-1 correlations in Fig. 8 are small. However, the spin-1/2 - 
spin-1/2 correlations in Fig. 7 
show an unexpected behavior for small $J_2$. We find that the spin-1/2's on 
even and odd sites appear to decouple into two separate chains, with the 
correlation being very small between spins belonging to different chains; 
within each chain, the correlations have an antiferromagnetic character. In 
other words, $<{\vec S}_{2,i} \cdot {\vec S}_{2,j}>$ is small if $i-j$ is 
odd, and it oscillates as $(-1)^{(i-j)/2}$ if $i-j$ is even. We call this the 
next-nearest-neighbor antiferromagnet (NNN-AFM). In the next section, we will 
provide some understanding of this behavior.

To understand better the nature of the changes in the ground state, we looked 
at the structure factors for the spin-1 - spin-1 and spin-1/2 - spin-1/2
correlations. These are respectively defined as
\bea
S^{11} (q) &=& \frac{1}{N} ~\sum_{i=1}^{N} ~<{\vec S}_{N+1} \cdot {\vec 
S}_{N+i}> ~\cos (qr_i) ~, \non \\
S^{22} (q) &=& \frac{4}{N} ~\sum_{i=1}^{N} ~<{\vec S}_{1} \cdot {\vec 
S}_{i}> ~\cos (qr_i) ~, 
\label{sf}
\eea
where we define 
\bea
r_i &=& i-1 \quad {\rm for} \quad 1 \le i \le \frac{N}{2} ~, \non \\
&=& N+1-i \quad {\rm for} \quad \frac{N}{2} + 1 \le i \le N ~,
\eea
to account for the periodic boundary conditions, and $q$ takes the values 
$2\pi n/N$, where $n=0,1,...,N-1$. We have included the factors of $1/S_i^2$ 
(equal to 1 and 4 for spin-1 and spin-1/2 respectively) on the right hand 
sides of Eq. (\ref{sf}) to make it easier to compare the magnitudes of 
$S^{11} (q)$ and $S^{22} (q)$. 

In Fig. 10, we show the values of $q$ where the two structure factors are 
maximum ($q_{max}$) as a function of $J_2$ for $N=12$. For $0 < J_2 \lesssim 
1$, $q_{max} = \pi /2$ for spin-1/2 and $\pi$ for spin-1. (The NNN-AFM 
behavior of the spin-1/2's discussed in the next section will explain why 
$q_{max} = \pi/2$ for spin-1/2 for small values of $J_2$). For $1.25 \lesssim 
J_2 \lesssim 1.75$, $q_{max} =0$ for spin-1/2 and $\pi$ for spin-1; this 
suggests that the ground state is in a canted state with a period of
two unit cells as shown in Fig. 1. 
For $1.9 \lesssim J_2 \lesssim 3.8$, $q_{max} = \pi /2$ for both spin-1/2 and 
spin-1; this suggests a spiral phase with a period of four unit cells. 
Finally, for $J_2 \gtrsim 3.8$, $q_{max}$ is equal to 0 for both spin-1/2 and 
spin-1; this is expected in the ferrimagnetic state. 

It is possible that the period two and period four states which are suggested
by the structure factor for $N=12$ (the periodic boundary conditions only
allow some limited periodicities for small systems) will turn into states with
longer periods (which change more smoothly with $J_2$) if we go to larger 
system sizes. The repeated level crossings between $k=0$ and $\pi$ shown in 
Table 1 also support this scenario \cite{tonegawa}.

Fig. 11 shows the values of $S^{ii} (q_{max})$ as a function of $J_2$. Once 
again, we see large fluctuations near $J_2 =1.1$, 1.9 and 3.8. The structure 
factors are relatively large for both large (ferrimagnetic) and small values 
of $J_2$, and is smaller for intermediate values of $J_2$.

Finally, we examined the possibility of dimerization, namely, whether the
ground state spontaneously breaks the invariance of the Hamiltonian under
translation by one unit cell. [The unit cell of our system has half-odd-integer
spin, and such systems are quite susceptible to dimerization in one dimension].
A simple way to study this question is to see if the spin-1/2 - spin-1/2 
correlations between site $1$ and its neighbors at sites $2$ and $N$ (i.e., 
$<{\vec S}_1 \cdot {\vec S}_2>$ and $<{\vec S}_1 \cdot {\vec S}_N>$) are equal.
The problem is that the energy eigenstates we have found are also eigenstates 
of momentum and are therefore translation invariant; hence the two correlations
will be equal in such states. However, if dimerization does occur,
we expect that the ground state (called $|1>$) will be almost degenerate 
with an excited state (called $|2>$) \cite{oshikawa}. Although each of these
would be eigenstates of momentum and therefore translation invariant,
the linear combinations $|2+> = (|1> + |2>)/{\sqrt 2}$ and $|2-> = (|1> - 
|2>)/{\sqrt 2}$ would not be translation invariant. Now, we see from Fig. 6
that the ground state is almost degenerate with the first excited state
(and both are singlets) at two values of $J_2$, namely, 1 and 2. We
therefore examine the two correlations mentioned above in the four states
$|1>$, $|2>$, $|2+>$ and $|2->$ at those two values of $J_2$. The results are
shown in Table 2. We see that the states $|2+>$ and $|2->$ do show an asymmetry 
in the two nearest neighbor correlations (and the values of the correlations 
are exchanged between the two states). However, the numerical values of
all the correlations are quite small, so there is no clear evidence for 
dimerization.

\vspace{0.4cm}

\begin{center}
\begin{tabular}{|c|c|c|c|} \hline
~~~~$J_2$~~~~ & ~~ State ~~ & ~~~$<{\vec S}_1 \cdot {\vec S}_2>$~~~ & 
~~~$<{\vec S}_1 \cdot {\vec S}_8>$~~~ \\ \hline
1 & $|1>$ & -0.00562 & -0.00561 \\
1 & $|2>$ & -0.07239 & -0.07240 \\
1 & $|2+>$ & -0.15152 & 0.07350 \\
1 & $|2->$ & 0.07351 & -0.15151 \\ \hline
2 & $|1>$ & 0.04198 & 0.04195 \\
2 & $|2>$ & 0.06867 & 0.06871 \\
2 & $|2+>$ & 0.09063 & 0.02004 \\
2 & $|2->$ & 0.02002 & 0.09062 \\ \hline
\end{tabular}
\end{center}
\vspace{0.2cm}

\noindent Table 2. The correlations of a spin-1/2 with its two neighboring
spin-1/2's in the ground state ($|1>$), first excited state ($|2>$), and 
the two linear combinations ($|2+>$ and $|2->$) at $J_2=$ 1 and 2, for 
$N=8$.
\vspace{0.4cm}

\section{Next-nearest-neighbor antiferromagnet near $J_2 =0$}

In this section, we will study the system for small values of $J_2$ using 
perturbation theory and the idea of an effective Hamiltonian. (A more 
detailed discussion of the ideas in this section is given in Ref. \cite{ravi}).
We write the Hamiltonian in Eq. (\ref{ham}) as the sum $H=H_0 +V$, where
\bea
H_0 &=& \sum_{i=N+1}^{2N} ~{\vec S}_i \cdot {\vec S}_{i+1} ~, \non \\
V &=& J_2 \sum_{i=1}^N ~{\vec S}_i \cdot [~{\vec S}_{i+N} + 
{\vec S}_{i+N+1})~]~.
\eea
For $J_2 =0$, we have an antiferromagnetic spin-1 chain with a coupling equal
to 1, and $N$ decoupled spin-1/2's. (Every state of the system will have a 
degeneracy of $2^N$ due to the decoupled spin-1/2's). It is known that the 
ground state of a spin-1 chain is a singlet with an energy $E^1_0 = -1.40148 
N$, and it is separated by a gap of $\Delta E^1 = 0.41050$
from the first excited state which is a triplet \cite{white1}. 

Let us denote the eigenstates of $H_0$ for the spin-1 chain by $|\psi^1_i>$
with energy $E^1_i$, where $i=0$ denotes the ground state. The states of the 
spin-1/2 sites will be denoted by $|\psi^{1/2}_j>$. The eigenstates of the full
Hamiltonian $H$ can therefore be written as linear combinations of the form
\beq
|\psi_a> ~=~ \sum_{i,j} ~c_{a,i,j} ~|\psi^1_i> \otimes |\psi^{1/2}_j> ~,
\eeq
where the $c_{a,i,j}$ are appropriate coefficients.

We will now expand up to second order in the perturbation $V$ to find an 
effective Hamiltonian $H_{eff}$ which acts within the subspace of the $2^N$ 
ground states which are degenerate for $J_2=0$. The Hamiltonian $H_{eff}$ will 
only act on the spin-1/2's. To first order in $V$, we have
\beq
H_{1,eff} ~=~ <\psi^1_0| ~V~ | \psi^1_0> ~.
\label{heff1}
\eeq
Since $V$ involves both spin-1/2 and spin-1 operators, and the expectation 
value in (\ref{heff1}) is being taken in a spin-1 state, we see that 
$H_{1,eff}$ will only involve spin-1/2 operators as desired. Now, the 
expectation value in (\ref{heff1}) is equal to zero, because $V$ is linear in
the spin-1 operators (which are not rotationally invariant), while 
$|\psi^1_0>$ is a singlet and is therefore rotationally invariant. 

We therefore have to go to second order in $V$. We then have
\beq
H_{2,eff} ~=~ \sum_{i \ne 0} ~\frac{<\psi^1_0| ~V~ | \psi^1_i> <\psi^1_i| ~
V~ | \psi^1_0>}{E^1_0 - E^1_i} ~,
\label{heff2}
\eeq
Clearly, this will be an operator which is of degree 2 or less in the spin-1/2
operators. Since the state $|\psi^1_0>$, the sum over states $\sum_{i\ne 0} 
|\psi^1_i> <\psi^1_i| /(E^1_0 - E^1_i)$ and $V$ are all invariant under 
rotations and translations, $H_{2,eff}$ must have the same invariances. The 
only operators which are of degree 2 or less in spin-1/2's and are rotationally
invariant are a constant and products of the form ${\vec S}_i \cdot {\vec 
S}_j$. Using translation invariance, we see that $H_{2,eff}$ must take the form
\bea
H_{2,eff} = & & N a ~+~ N J_2^2 b \non \\
& & + ~J_2^2 ~ \sum_i ~[~ c_1 {\vec S}_i \cdot {\vec S}_{i+1} ~+~
c_2 {\vec S}_i \cdot {\vec S}_{i+2} ~+~ c_3 {\vec S}_i \cdot {\vec S}_{i+3} ~
+~ \cdot \cdot \cdot ~]~ ,
\label{heff3}
\eea
where $a,b,c_1,c_2,...$ are numbers which are independent of $J_2$, 
and appropriate periodic
boundary conditions are assumed in the summations over $i$. For a periodic
system with $N$ spin-1/2's, the subscript $i$ of $c_i$ goes from 1 to $N/2$
(since $N$ is even), so a total of $2+N/2$ numbers have to determined. 
These numbers will of course depend on the system size; but since the
ground state of a spin-1 chain has a finite correlation length, we would
expect these numbers to converge quickly to some values as $N$ becomes large.
(We will assume that $J_2$ is small enough so that terms of order 
$J_2^3$ and higher can be neglected in comparison with the terms of order
$J_2^2$ which we are interested in).

A direct computation of the constants $a,b,c_i$ in Eq. (\ref{heff3}) using the
expression in (\ref{heff2}) is difficult because we would need to accurately 
determine all the energy levels and eigenstates of a spin-1 chain as well as
all the matrix elements appearing in that expression. We therefore
assume the form in Eq. (\ref{heff3}) (which we have so far found purely on 
grounds of symmetry), and numerically determine the constants as follows. To 
determine the first number $a$ in (\ref{heff3}), we set $J_2=0$ and 
numerically find the ground state energy which is equal to $Na$. Next, we 
turn on the $J_2$ couplings on the bonds connecting only two of the spin-1/2's,
say at sites $1$ and $n+1$, to the spin-1's. In other words, we set four of the
spin-1/2 - spin-1 couplings equal to $J_2$, and keep all the other spin-1/2 - 
spin-1 couplings equal to zero; let us call this truncated perturbation 
$V_1 + V_{n+1}$ (thus, $V=\sum_i V_i$). We ignore the $N-2$ spin-1/2's 
which are not coupled to the spin-1's. The energy levels of the system 
consisting of the spin-1 chain and two spin-1/2's will have four low-lying 
states (which would be degenerate with an energy of $Na$ if all the 
$J_2$'s had been set equal to zero). These four states are described by an 
effective Hamiltonian involving the two spin-1/2's of the form 
\beq
H_{ij,eff} ~=~ Na ~+~ J_2^2 ~(~ 2b ~+~ c_n {\vec S}_1 \cdot
{\vec S}_{n+1} ~)~.
\label{heff4}
\eeq
The important point is that the constants $b$ and $c_n$ in this expression
have the same values as in Eq. (\ref{heff3}) where all the $J_2$ couplings
are turned on. The reason for this can be
traced back to the expression in (\ref{heff2}) which can be used for either 
the full perturbation $V$ or the truncated perturbation $V_1 + V_{n+1}$. 
A comparison between the two second order expressions shows that the constant
$b$ arises from the product of a spin-1/2 operator at site $1$ with itself
(when we take the product of the
two matrix elements in (\ref{heff2})); that is why it appears with
a factor of $N$ in (\ref{heff3}) and a factor of $2$ in (\ref{heff4}). On
the other hand, the constant $c_n$ comes from a product of a spin-1/2 operator
at site $1$ with a spin-1/2 operator at site $n+1$, and it comes
with the same factor in (\ref{heff3}) and (\ref{heff4}).

We can numerically
determine the constants $b$ and $c_n$ from the energies of the four 
low-lying states of the spin-1 chain plus two spin-1/2's; three of these
states will form a triplet with the same energy and one will form a singlet,
so that there will be only two equations in two unknowns. We can then repeat 
the procedure and determine all the constants $c_i$ by successively coupling 
various pairs of spin-1/2's to the spin-1 chain; in each case, we only have 
to look at the four low-lying energy levels to find $b$ and $c_i$. (The values
of $b$ that we get in the different cases should of course be consistent with
each other). This procedure will work provided that $J_2$ is small enough 
that the four-low lying energy levels lie far below the gap $\Delta E^1$ of 
the pure spin-1 chain, and the terms of third and higher orders are 
much smaller than those of second order. On the other hand, if we choose
$J_2$ to be too small, the energy splittings of $J_2^2$ are very
small, and the determination of the constants $b$ and $c_i$ will suffer from
large numerical uncertainties. For our calculations with $N=8$, we found that
taking $J_2 =0.1$ gives reasonably accurate and self-consistent results. We 
found the following values of the six numbers:
\bea
a &=& -1.41712 ~, \quad \quad b ~= -0.12665 ~, \non \\
c_1 &=& 0.0183 ~, \quad \quad ~~~~c_2 = 0.1291 ~, \non \\
c_3 &=& -0.0108 ~, \quad \quad ~~c_4 = 0.0942 ~.
\label{num}
\eea
We see that the value of $a$ found for $N=8$ agrees quite well with the 
thermodynamic value ($N \rightarrow \infty$) of $-1.40148$ quoted earlier 
\cite{white1}.

Looking at the values of $c_i$ in (\ref{num}), we observe the curious pattern 
that $c_2$ is the largest number, followed by $c_4$; the numbers $c_1$ and
$c_3$ are much smaller in comparison. Thus the spin-1/2's governed by the
effective Hamiltonian in (\ref{heff3}) seem to break up into two chains, one
consisting of the odd numbered sites, and the other with the even numbered 
sites. Each of the chains has a nearest neighbor coupling of $c_2 J_2^2$ 
which is antiferromagnetic; we therefore call this the NNN-AFM. 
This explains the numerical result that the structure factor of the spin-1/2's
is peaked at $q=\pi /2$ and that the next-nearest-neighbor correlation is the
largest in magnitude (and has a negative sign) for small $J_2$. 

Note, however, that the next-nearest-neighbor coupling in each chain 
(proportional to $c_4$ which is about 0.73 times $c_2$) is also 
antiferromagnetic and is not much smaller than the nearest neighbor
coupling, so each of the spin-1/2 chains is strongly frustrated. For such a 
strong frustration, it is known that a spin-1/2 chain is disordered 
with a small correlation length of about two lattice spacings (this implies a 
correlation length of about four lattice spacings in the sawtooth system), 
and is also strongly dimerized \cite{white2}. The small correlation length 
is supported by the correlation data for $N=12$ and $J_2 =0.1$; we find that 
the ratio of spin-1/2 - spin-1/2 correlations $<{\vec S}_1 \cdot {\vec S}_5> /
<{\vec S}_1 \cdot {\vec S}_3> \simeq - 0.411$, while the ratio of spin-1 - 
spin-1 correlations $<{\vec S}_{13} \cdot {\vec S}_{15}> / <{\vec S}_{13} 
\cdot {\vec S}_{14}> \simeq - 0.552$. Thus the spin-1/2 correlations (within 
each chain) decay faster with increasing distance than the spin-1 
correlations (which have a correlation length of six lattice spacings).

To examine the possibility of dimerization, we use a method similar
to the one used at the end of Sec. III to look for dimerization at $J_2=1$ 
and 2. However, the present case is different for the following reasons.
First, we are now considering a NNN-AFM, so we have to
check if the spin-1/2 - spin-1/2 correlations between a site and its 
next-nearest-neighbors are equal. Secondly, we have to simultaneously look for
dimerization in the two spin-1/2 chains which are almost decoupled from
each other. If there is dimerization, we expect four low-lying states which
are almost degenerate with each other. For $N=8$, these four states will
exhibit dimerization in the four quantities $<{\vec S}_1 \cdot {\vec S}_3>$, 
$<{\vec S}_1 \cdot {\vec S}_7>$, $<{\vec S}_2 \cdot {\vec S}_4>$, and 
$<{\vec S}_2 \cdot {\vec S}_8>$. For $J_2 =0.1$, we find that there is a
non-degenerate ground state $|1>$, and two degenerate excited states ($|2>$
and $|3>$) which are separated from the ground state by a small gap of 
$0.000674$. (The next excited state, $|4>$, is separated from the ground state
by a gap of 0.001277; for simplicity, we will not include this state in the
following computations). The states $|1>$, $|2>$ and $|3>$ are all translation 
invariant, and therefore cannot show dimerization. We therefore consider the 
four linear combinations, $|2\pm> = (|1> \pm |2>)/{\sqrt 2}$ and 
$|3\pm> = (|1> \pm |3>)/{\sqrt 2}$ which are not translation invariant. We 
then compute the four correlations mentioned above in all the seven states; 
the results are shown in Table 3. We observe a substantial amount of 
dimerization in the states $|2\pm>$ and $|3\pm>$. If we define the 
dimerization in the two chains to be \cite{white2}
\bea
d_1 &=& <{\vec S}_1 \cdot {\vec S}_3> ~-~ <{\vec S}_1 \cdot {\vec S}_7> ~, 
\non \\
d_2 &=& <{\vec S}_2 \cdot {\vec S}_4> ~-~ <{\vec S}_2 \cdot {\vec S}_8> ~,
\label{dimer}
\eea
we see that the dimerizations in states $|2\pm>$ and $|3\pm>$ are both equal 
to about $\pm 0.6085$. Further, the correlations in these four states show all
the four possible patterns of dimerization which can occur for two chains.

\vspace{0.4cm}

\begin{center}
\begin{tabular}{|c|c|c|c|c|} \hline
~~ State ~~ & ~~~$<{\vec S}_1 \cdot {\vec S}_3>$~~~ & 
~~~$<{\vec S}_1 \cdot {\vec S}_7>$~~~ &
~~~$<{\vec S}_2 \cdot {\vec S}_4>$~~~ &
~~~$<{\vec S}_2 \cdot {\vec S}_8>$~~~ \\ \hline
$|1>$ & -0.49765 & -0.49765 & -0.49765 & -0.49765 \\
$|2>$ & -0.24630 & -0.24629 & -0.24630 & -0.24631 \\
$|3>$ &  -0.24639 & -0.24638 & -0.24625 & -0.24624 \\
$|2+>$ & -0.06774 & -0.67620 & -0.67620 & -0.06775 \\
$|2->$ & -0.67621 & -0.06774 & -0.06775 & -0.67621 \\
$|3+>$ & -0.06781 & -0.67622 & -0.06765 & -0.67624 \\
$|3->$ & -0.67623 & -0.06781 & -0.67625 & -0.06765 \\
\hline
\end{tabular}
\end{center}
\vspace{0.2cm}

\noindent Table 3. The correlations of the spin-1/2's at sites 1 and 2 with 
their two next-nearest-neighboring spin-1/2's in the ground state ($|1>$), 
first excited states ($|2>$ and $|3>$), and the four linear combinations 
($|2 \pm>$ and $|3 \pm>$) at $J_2= 0.1$, for $N=8$.
\vspace{0.4cm}

The occurrence of a NNN-AFM with strong frustration
for small values of $J_2$ is one of the interesting features of the 
spin-1/2 - spin-1 sawtooth chain. Although the spin-1 chain is gapped and 
therefore plays no direct role at energy scales much smaller than
$J_1 =1$, it perturbatively induces an unusual kind of interaction between 
the spin-1/2's which leads to a non-trivial behavior for that subsystem.

\section{Macroscopic magnetization jump at $J_2 =2$}

In this section, we will discuss the phenomenon of a macroscopic
magnetization jump which occurs in the sawtooth chain for arbitrary values of 
$S_1$ and $S_2$ if $J_2=2$. In general, this phenomenon can occur in 
highly frustrated quantum antiferromagnets in which one of the spin wave 
modes (above the fully polarized ferromagnetic state)
is completely dispersionless. When a uniform magnetic field is applied
to the system, the magnetization can show a macroscopic jump at the saturation
field $h_s$ (defined as the minimum field for which all the spins are aligned
in the ground state) \cite{schulenburg,schnack}. By macroscopic we mean that 
the magnetization per unit cell jumps by a finite amount $\Delta m$ at $h=h_s$.
This occurs if, (i) there is a special kind of ferromagnetic one-magnon 
eigenstate of the Hamiltonian which is spatially localized (a few lattice 
spacings), (ii) this eigenstate has the lowest energy amongst all the 
one-magnon eigenstates, (iii) the energy of this one-magnon state is negative 
with respect to the fully aligned state if $h < h_s$, and (iv) there are no 
multi-magnon bound states 
with energy lower than the sum of the individual one-magnon states. If all 
these conditions are satisfied, then for a certain range of values of the 
magnetic field below $h_s$, the lowest energy state is one in which there is 
a macroscopic number of these magnons localized in disjoint regions of the 
lattice. Eventually, as the field $h$ is increased, the energy of these magnons
will cross zero at $h=h_s$ and then turn positive; for $h>h_s$, therefore, the
lowest energy state will be the one in which all the spins are aligned with 
the field. Hence there will be a macroscopic magnetization jump at $h_s$.

For the sawtooth chain with spins $S_1$ and $S_2$, we consider a Hamiltonian
which is the sum of the one given in Eq. (\ref{ham}) and a magnetic
field term given by
\beq
H_{mag} ~=~ - h ~\sum_{i=1}^{2N} ~S_{i,z} ~,
\eeq
where we have assumed the same value of the gyromagnetic ratio $g$ for spins
$S_1$ and $S_2$, and we have absorbed $g$ in the definition of the magnetic 
field $h$. For this system with $J_2 =2$, the special one-magnon state 
(above the ferromagnetic state) is 
a superposition of three states: $|2,n-1>$ in which the spin-$S_2$
in triangle $n-1$ has $S_z = S_2 -1$ (and all the other spins have the
maximum possibles values of $S_z$), $|1,n>$ in which the spin-$S_1$
in triangle $n$ has $S_z = S_1 -1$, and $|2,n>$ in which the spin-$S_2$
in triangle $n$ has $S_z = S_2 -1$. The particular superposition of these
three states which is an eigenstate of the total Hamiltonian is given by
\beq
|\psi_n> ~=~ |2,n-1> ~+~ |2,n> ~-~ 2 {\sqrt {\frac{S_2}{S_1}}} ~|1,n> ~.
\label{psin}
\eeq
The energy of this state with respect to the fully aligned state is given
by $E= h - 4 (S_1 + S_2)$. (The total spin of this state is $N(S_1+S_2)-1$,
since it has total $S_z = N(S_1 + S_2)-1$ and is annihilated by total $S_+$).
If we look at all the one-magnon states (above the ferromagnetic state), we 
find that they have two branches with the dispersions $\omega_- = h - 4 (S_1 +
S_2)$ (which is independent of the momentum and is equal to the energy of the 
localized one-magnon state $|\psi_n>$), and $\omega_+ = h - 4 S_1 
\sin^2 (k/2)$ which is greater than $\omega_-$ for all values of $k$. 

We thus see that the state $|\psi_n>$ meets the conditions (i) and (ii) given 
above, and its energy is lower than that of the fully aligned state if 
$h<h_s$, where
\beq
h_s ~=~ 4 (S_1 + S_2) ~.
\eeq
We therefore identify $h_s$ as the saturation field, and we expect a 
macroscopic jump in the magnetization when $h$ crosses $h_s$ 

The magnitude of the magnetization jump can be found as follows. Since each
of the special one-magnon states involves three sites, at most $N/2$ 
such states can exist in disconnected regions of a chain with $N$ triangles.
The lowest energy of a state with $n$ magnons will be less than the energy of 
the fully aligned state by an amount equal to $n[h - 4 (S_1 +S_2)]$ as 
long as $n \le N/2$. Once the number of magnons exceeds $N/2$, some of them 
will be close enough to interact (repulsively) with each other, and we no 
longer expect the energy to vary linearly with the number of magnons.
Hence, when the magnetic field is lowered slightly below $h_s$, we
expect the magnetization to abruptly drop from the maximum possible value of
$M_{max} = N(S_1 + S_2)$ to $M_{max} - N/2$. The magnetization jump is
therefore given by $\Delta M = N/2$. The ratio $\Delta M /M_{max} = 
1/2(S_1 +S_2)$ goes to zero in the classical limit $S_1 , S_2 
\rightarrow \infty$. The magnetization jump is therefore a true quantum
effect as emphasized in Ref. \cite{schulenburg}.

For general values of $S_1$ and $S_2$, we have not analytically checked 
condition (iv) that there are no multi-magnon bound states with energy lower 
than the sum of one-magnon bound states. However, this is numerically found 
to be true in many models due to the absence of attractive interactions 
between the magnons \cite{schulenburg,schnack}. This is also found to be true 
in our system with $S_1 =1$ and $S_2 =1/2$, as the data given below shows.

For $N=12$, we numerically find that in the absence of a magnetic field,
the lowest energy $E_0$ in subspaces with different values of the total 
$S_z$ is given by, $E_0 (S_z =18) = 36$, $E_0 (S_z =17) = 30$, $E_0 (S_z =16)
= 24$, ..., $E_0 (S_z =12) = 0$, and $E_0 (S_z =11) = -5.167392$. 
Thus, when the magnetic field strength is lowered just below $h_s =6$, the
magnetization jumps abruptly from 18 to 12 in accordance with the arguments
given above.

\section{Discussion}

We have studied the ground state and low-energy properties of a spin-1/2 -
spin-1 sawtooth chain using SWT and exact diagonalization of finite systems.
Linear SWT shows that there are two phases (the ground state being 
ferrimagnetic in one phase and a singlet in the other phase) separated by the 
value of $J_2 =4$. In addition, $J_2 =2$ 
is special because all the classically degenerate states have total spin equal
to zero at that point, and $J_1 =1$ is special because the total spin
in each triangle is zero in all the classical ground states. 

The numerical studies indicate that there are four distinct regions. 
For $J_2 \gtrsim 3.8$, the ground state is ferrimagnetic, while for $J_2 
\lesssim 3.8$, it is a singlet. 
The structure factors suggest that the ground state is in a spiral state 
with a period of four unit cells for $1.9 \lesssim J_2 \lesssim 3.8$, and in 
a canted state with a period of two unit cells for $1.1 \lesssim 
J_2 \lesssim 3.8$. Near $J_2 =1$ and $2$, the gap between the ground state and
the first excited state is particularly small, and there are repeated level 
crossings, possibly indicating crossovers between ground states with 
different kinds of short-range correlations. (Numerical calculations on 
larger system sizes would be very useful for a complete understanding of the 
nature of the ground state for $1 \lesssim J_2 \lesssim 3.8$). Finally, the 
spin-1/2's form an interesting system called a NNN-AFM for $J_2 \lesssim 1$;
the ground state of this system has a short correlation length and is 
strongly dimerized. 

\newpage
\centerline{\bf Acknowledgments}
\vskip 0.5 true cm

DS thanks the Department of Science and Technology, India for financial 
support through grant no. SP/S2/M-11/00. NBI and JR thank the Deutsche 
Forschungsgemeinschaft for financial support (projects no. 436BUL/17/5/03 
and no. Ri 615/6-1). JR is indebted to J. Schulenburg for numerical assistance.

\newpage

\noindent {\bf Figure Captions}
\vskip .5 true cm

\noindent {1.} Picture of a sawtooth chain with 8 triangles indicating the site
labels for the spin-$S_2$'s at the vertices and the spin-$S_1$'s at the bases 
of the triangles, and the couplings $J_1$ and $J_2$. For the numerical studies,
we take $S_1 =1$ and $S_2 =1/2$. The arrows and angles ($\theta$) indicate a 
canted state in which all the spin-$S_2$'s are aligned with each other.

\noindent {2.} Spin wave dispersions in the ferrimagnetic phase for $J_2 = 5$,
$S_1 =1$ and $S_2 = 0.5$.

\noindent {3.} Non-vanishing spin wave dispersion in the singlet phase for
$J_2 = 2$, $S_1 =1$ and $S_2 = 0.5$.

\noindent {4.} Ground state energy as a function of $J_2$. The solid line 
shows the numerical data from exact diagonalization for a chain with 8 
triangles, while the dashed line shows the spin wave results. In the inset, 
the solid lines show piecewise linear fits to the numerical data to the left 
and right of $J_2 =3.8$, while the dotted lines show the continuations of the 
same two straight lines to the right and left of $J_2 =3.8$ respectively. 
This shows a small discontinuity in the slope at $J_2 \simeq 3.8$. 

\noindent {5.} Total spin of the first excited state as a function of $J_2$ 
for a chain with 8 triangles.

\noindent {6.} Energy gaps between the ground state and the first excited state
(lower curve) and the first non-singlet state (upper curve)
as a function of $J_2$ for a chain with 8 triangles. The ground state is a
singlet for the range of $J_2$ shown in the figure.

\noindent {7.} The spin-1/2 - spin-1/2 correlations as functions of $J_2$ for 
a chain with 12 triangles.

\noindent {8.} The spin-1/2 - spin-1 correlations as functions of 
$J_2$ for a chain with 12 triangles.

\noindent {9.} The spin-1 - spin-1 correlations as functions of $J_2$ for a 
chain with 12 triangles.

\noindent {10.} Values of $q$ where the structure factors $S^{11}(q)$ and
$S^{22}(q)$ are maximum as functions of $J_2$ for a chain with 12 triangles.

\noindent {11.} The structure factors $S^{11} (q_{max})$ and $S^{22} 
(q_{max})$ as functions of $J_2$ for a chain with 12 triangles.

\newpage

\begin{figure}[ht]
\begin{center}
\hspace*{-2cm}
\epsfig{figure=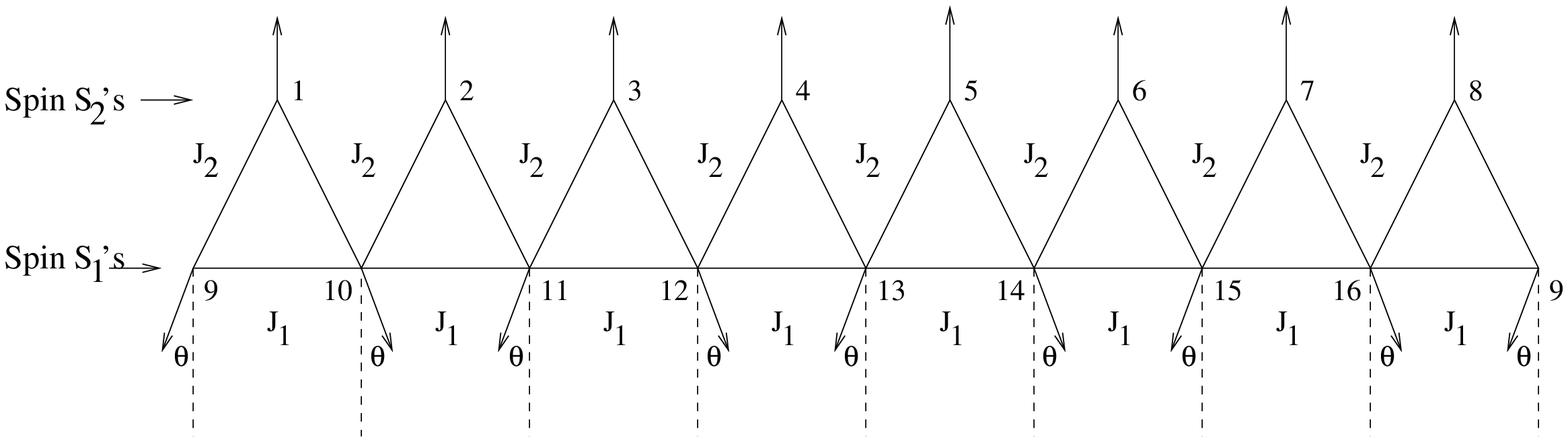,width=16cm}
\end{center}
\vspace*{2cm}
\centerline{Fig. 1}
\end{figure}

\newpage

\begin{figure}[ht]
\begin{center}
\hspace*{-2cm}
\epsfig{figure=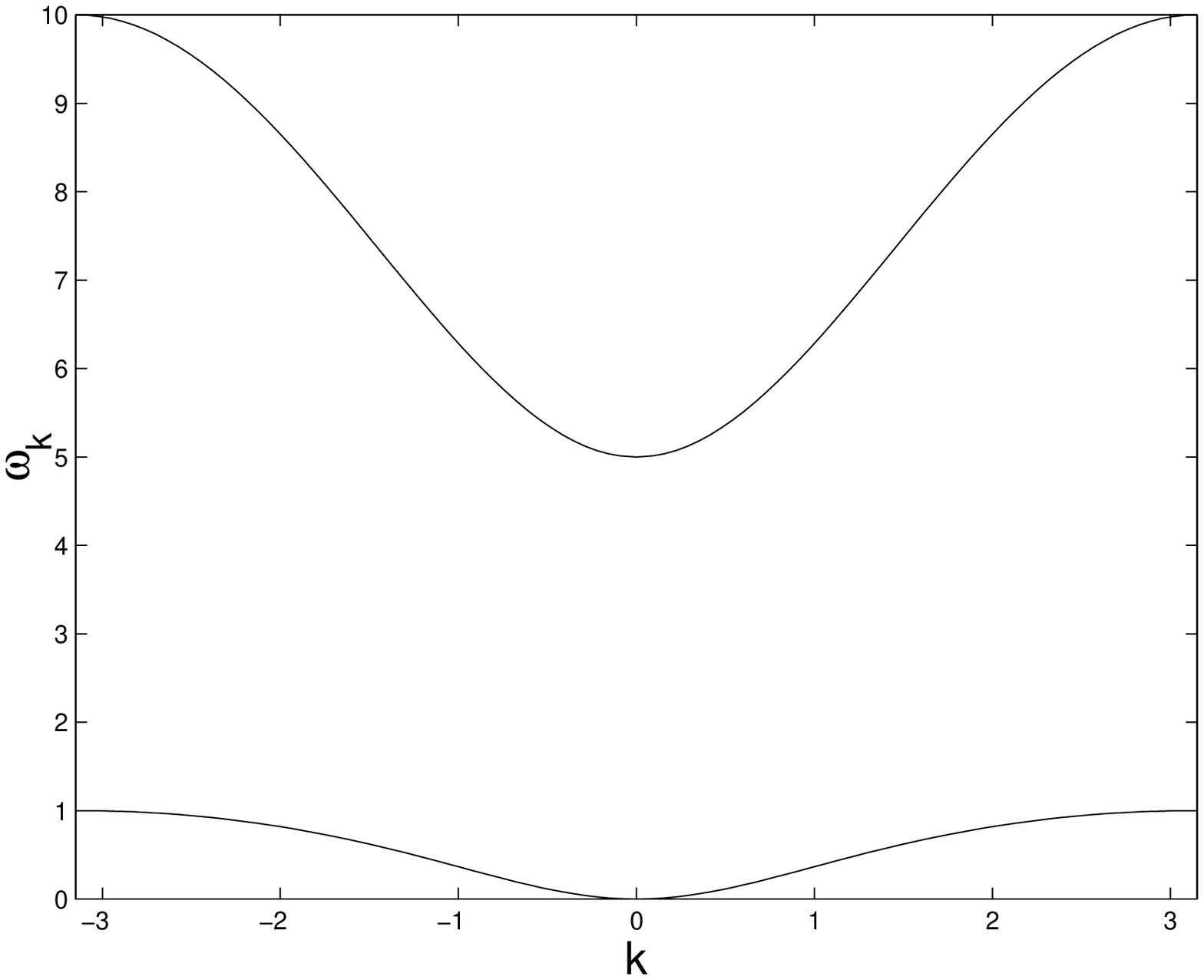,width=14cm}
\end{center}
\vspace*{2cm}
\centerline{Fig. 2}
\end{figure}

\newpage

\begin{figure}[ht]
\begin{center}
\hspace*{-2cm}
\epsfig{figure=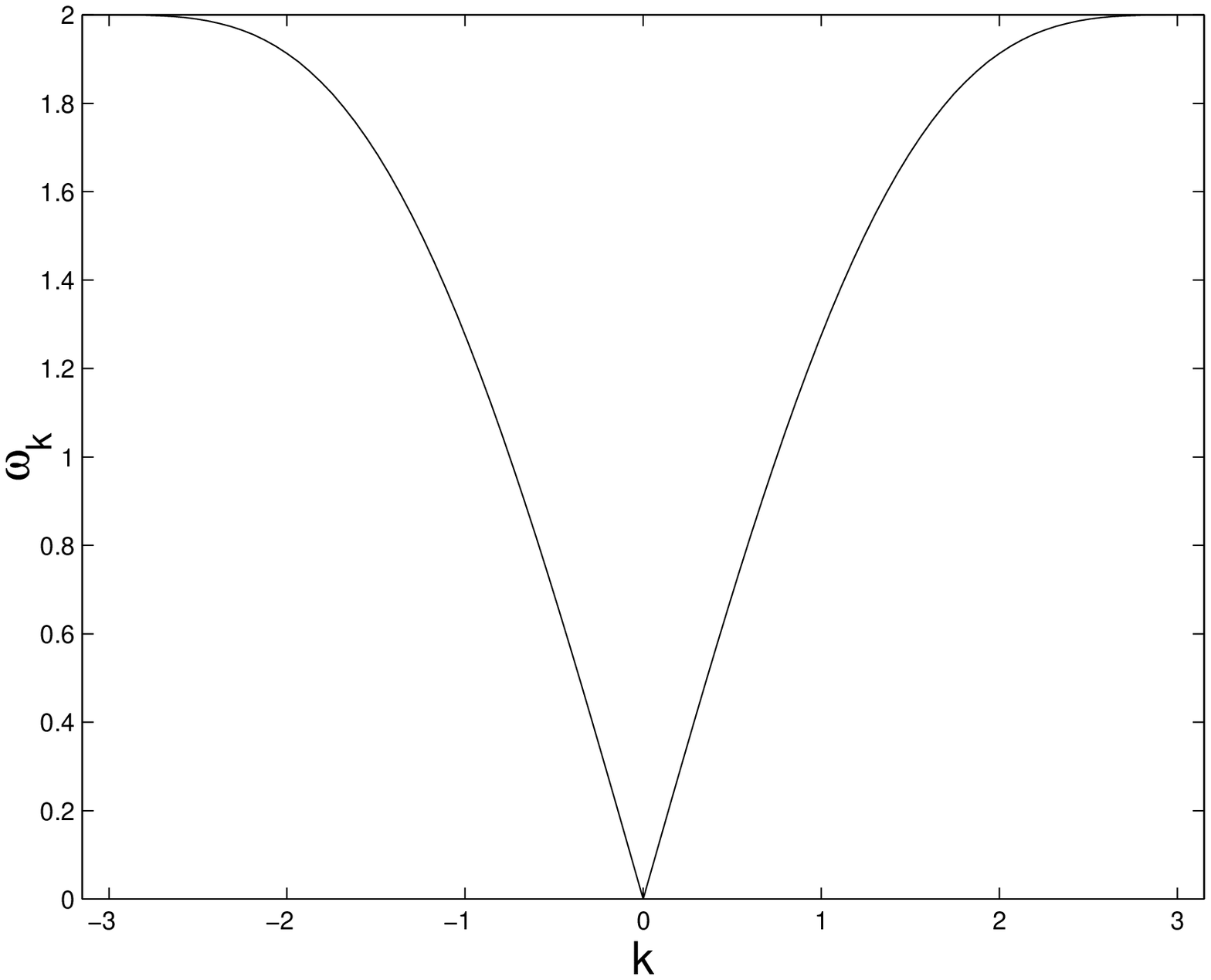,width=14cm}
\end{center}
\vspace*{2cm}
\centerline{Fig. 3}
\end{figure}

\newpage

\begin{figure}[ht]
\begin{center}
\hspace*{-2cm}
\epsfig{figure=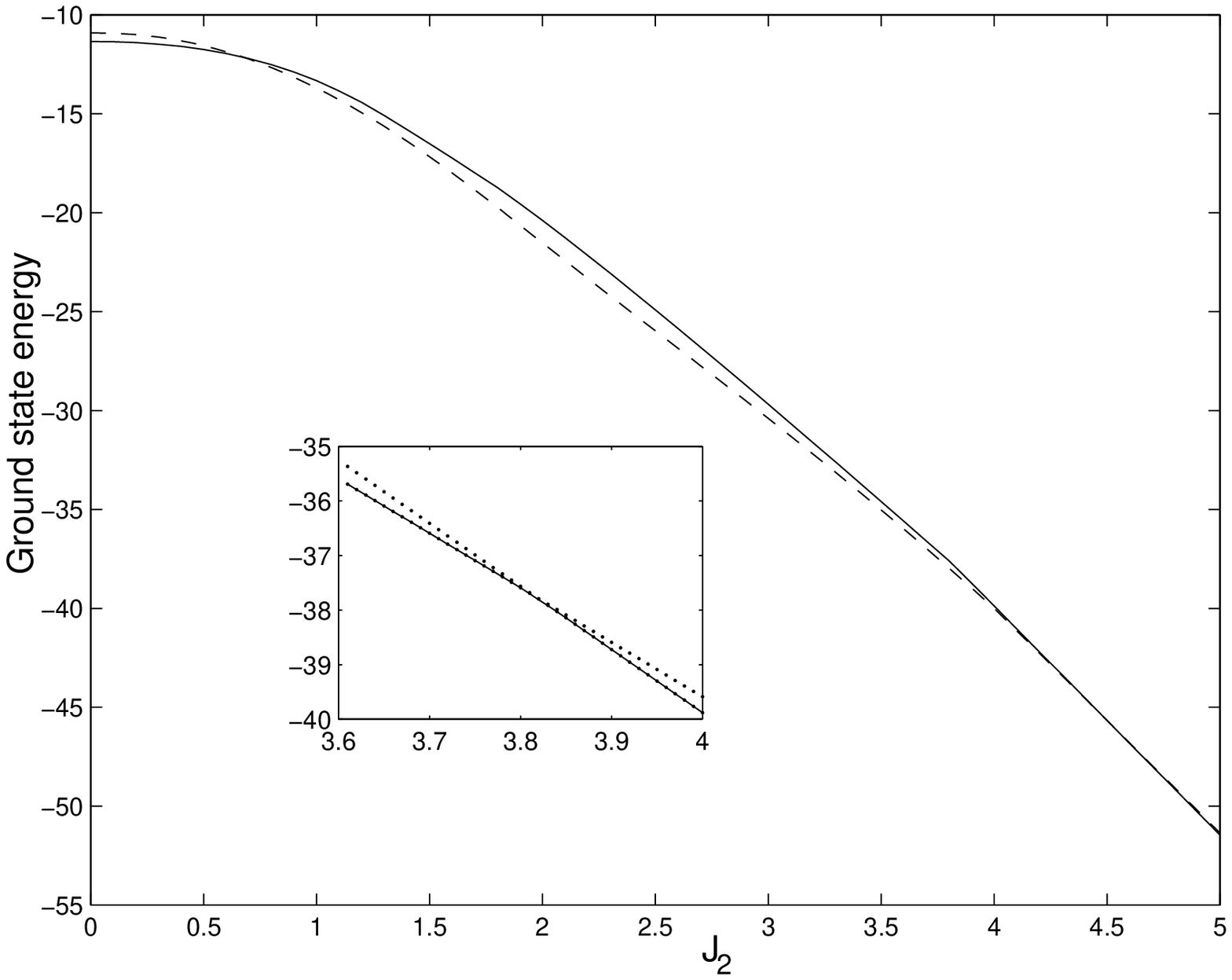,width=14cm}
\end{center}
\vspace*{2cm}
\centerline{Fig. 4}
\end{figure}

\newpage

\begin{figure}[ht]
\begin{center}
\hspace*{-2cm}
\epsfig{figure=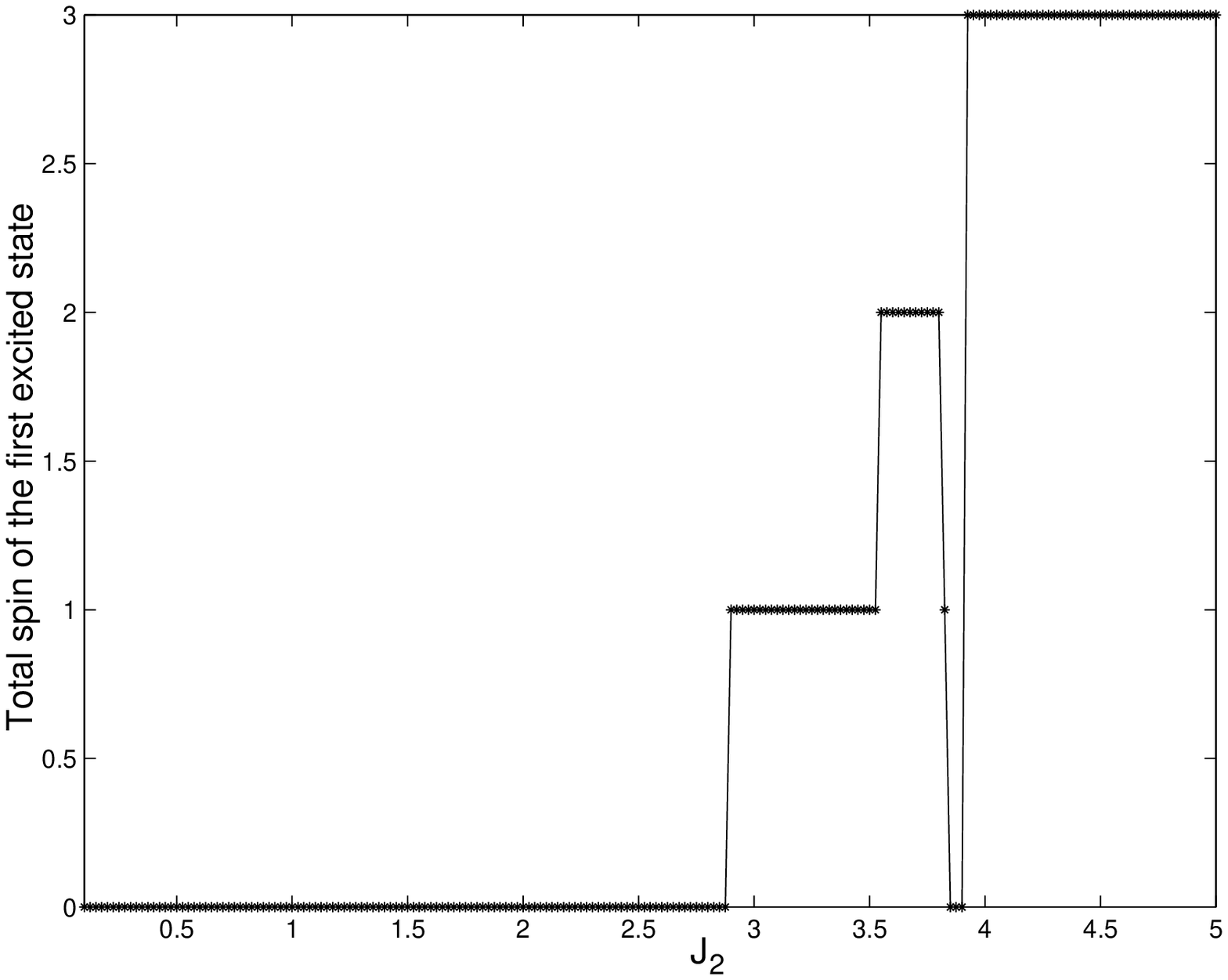,width=14cm}
\end{center}
\vspace*{2cm}
\centerline{Fig. 5}
\end{figure}

\newpage

\begin{figure}[ht]
\begin{center}
\hspace*{-2cm}
\epsfig{figure=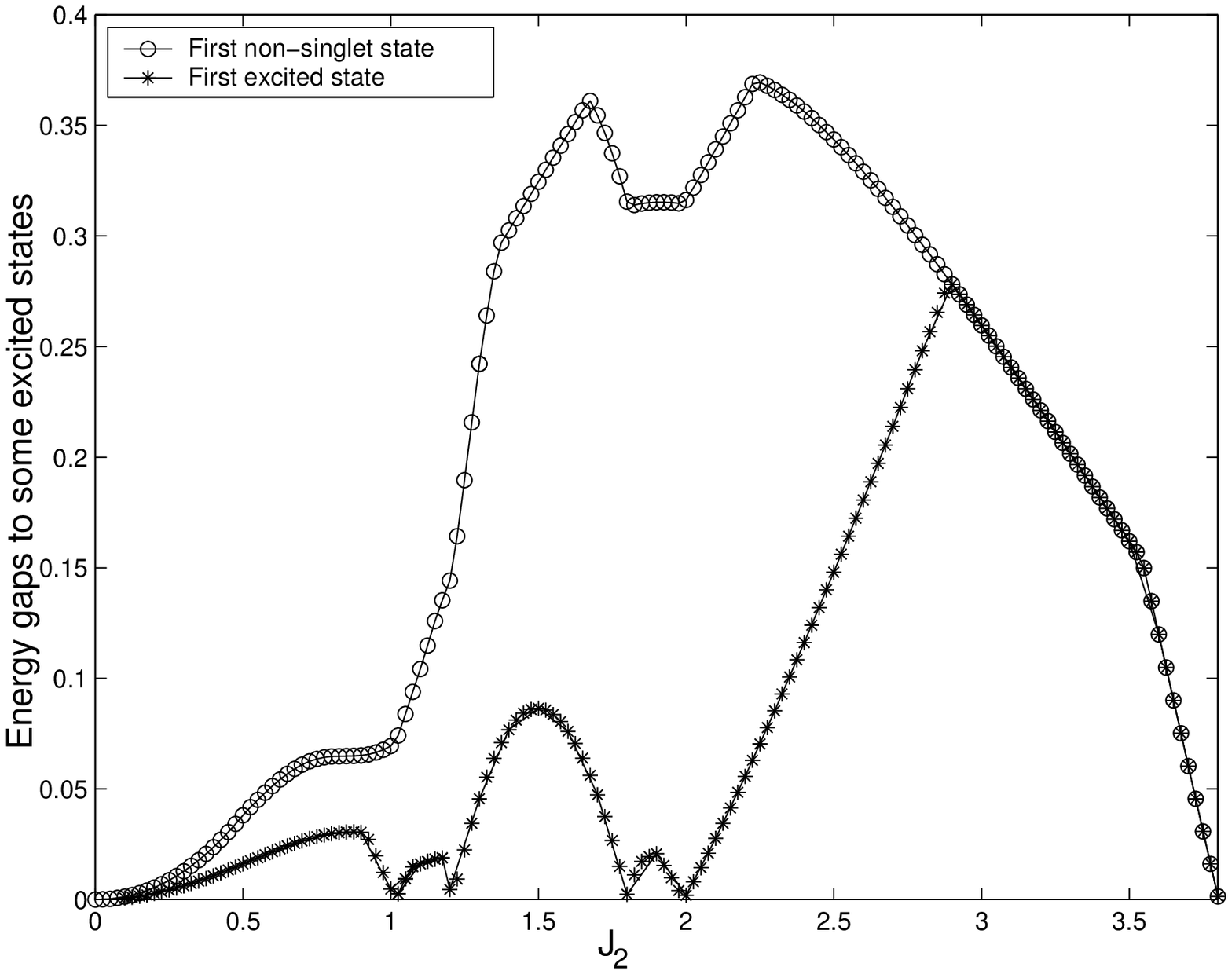,width=14cm}
\end{center}
\vspace*{2cm}
\centerline{Fig. 6}
\end{figure}

\newpage

\begin{figure}[ht]
\begin{center}
\hspace*{-2cm}
\epsfig{figure=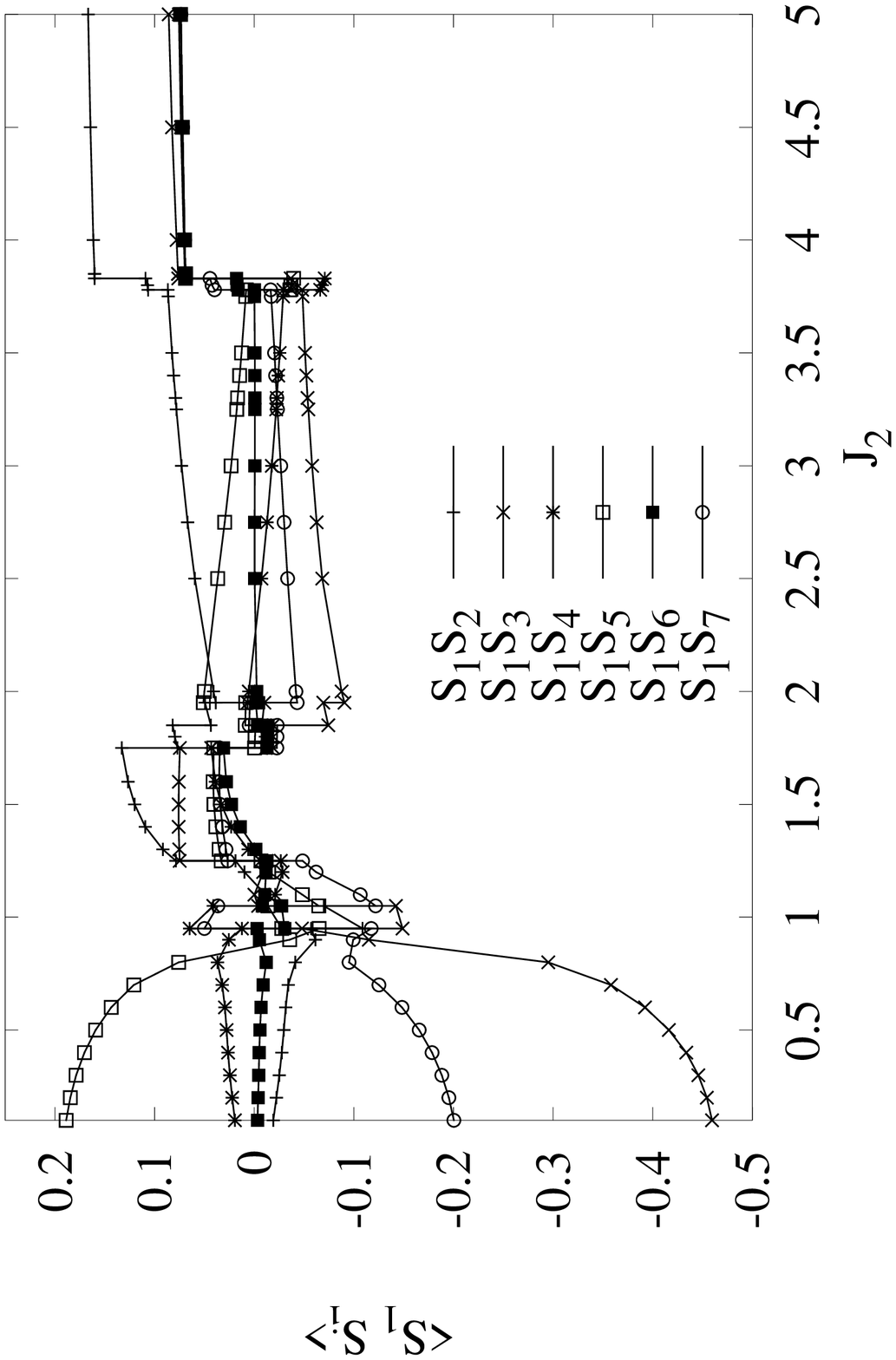,width=13cm}
\end{center}
\end{figure}
\centerline{Fig. 7}

\newpage

\begin{figure}[ht]
\begin{center}
\hspace*{-2cm}
\epsfig{figure=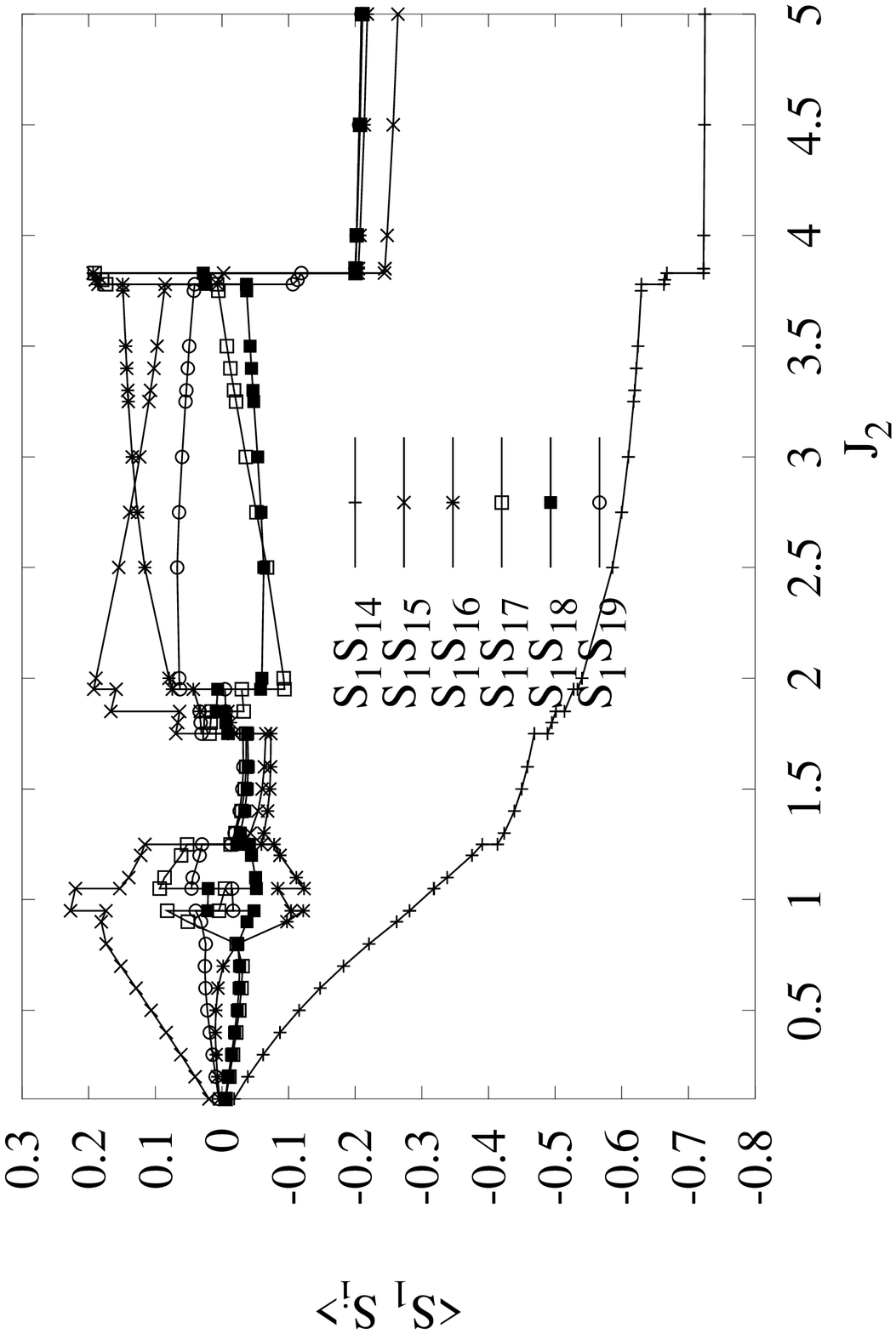,width=13cm}
\end{center}
\end{figure}
\centerline{Fig. 8}

\newpage

\begin{figure}[ht]
\begin{center}
\hspace*{-2cm}
\epsfig{figure=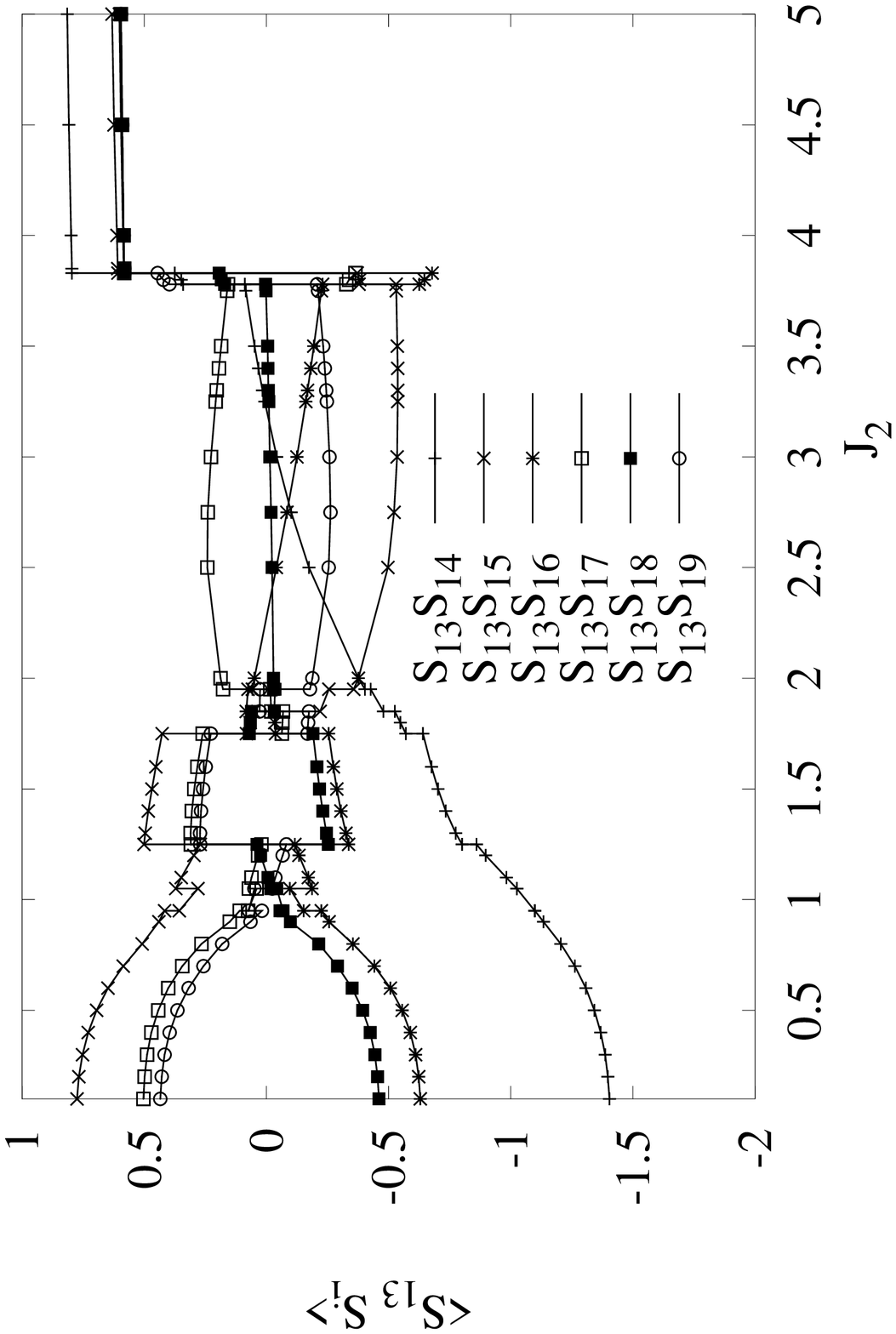,width=13cm}
\end{center}
\end{figure}
\centerline{Fig. 9}

\newpage

\begin{figure}[ht]
\begin{center}
\hspace*{-2cm}
\epsfig{figure=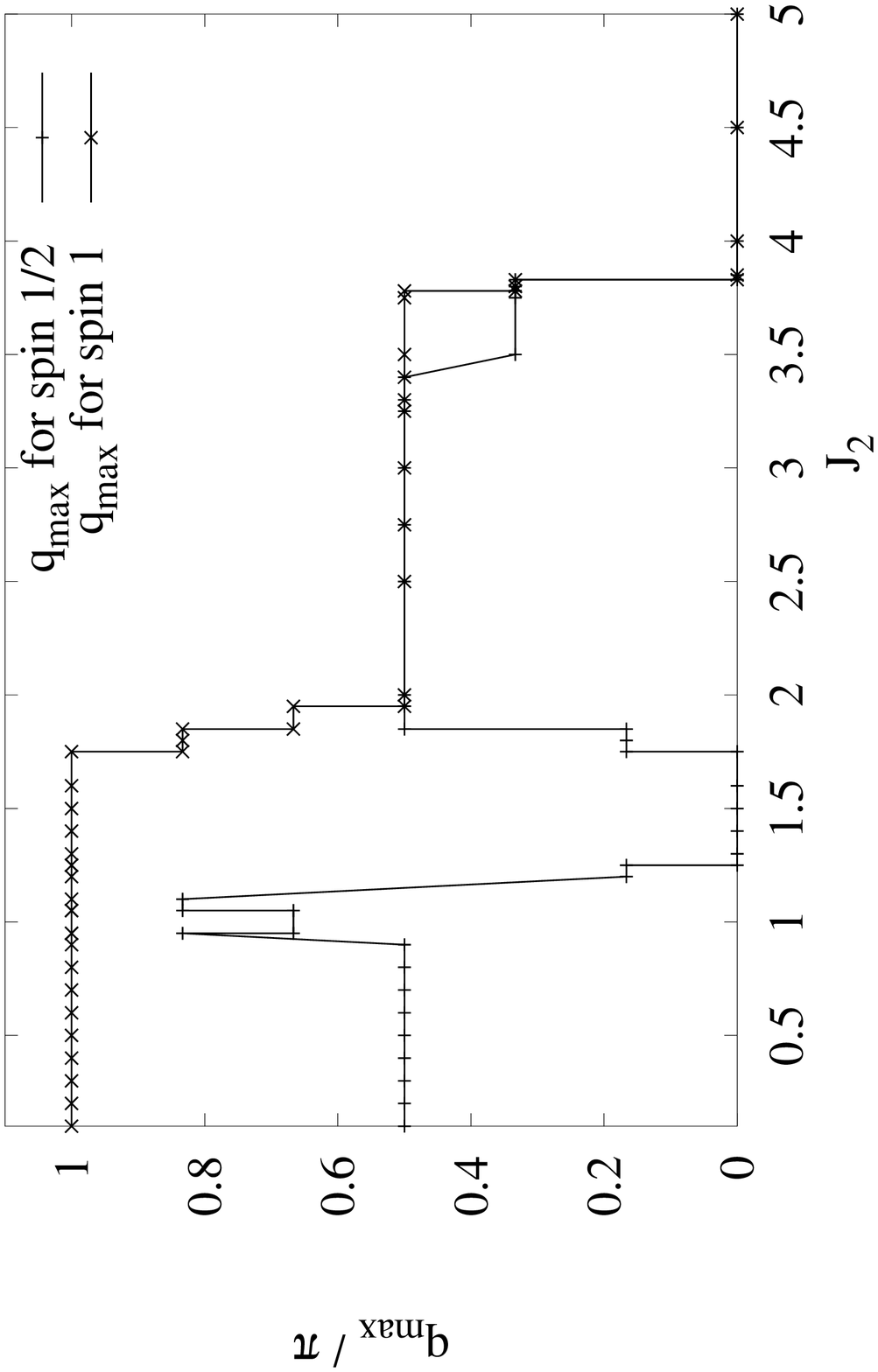,width=13cm}
\end{center}
\end{figure}
\centerline{Fig. 10}

\newpage

\begin{figure}[ht]
\begin{center}
\hspace*{-2cm}
\epsfig{figure=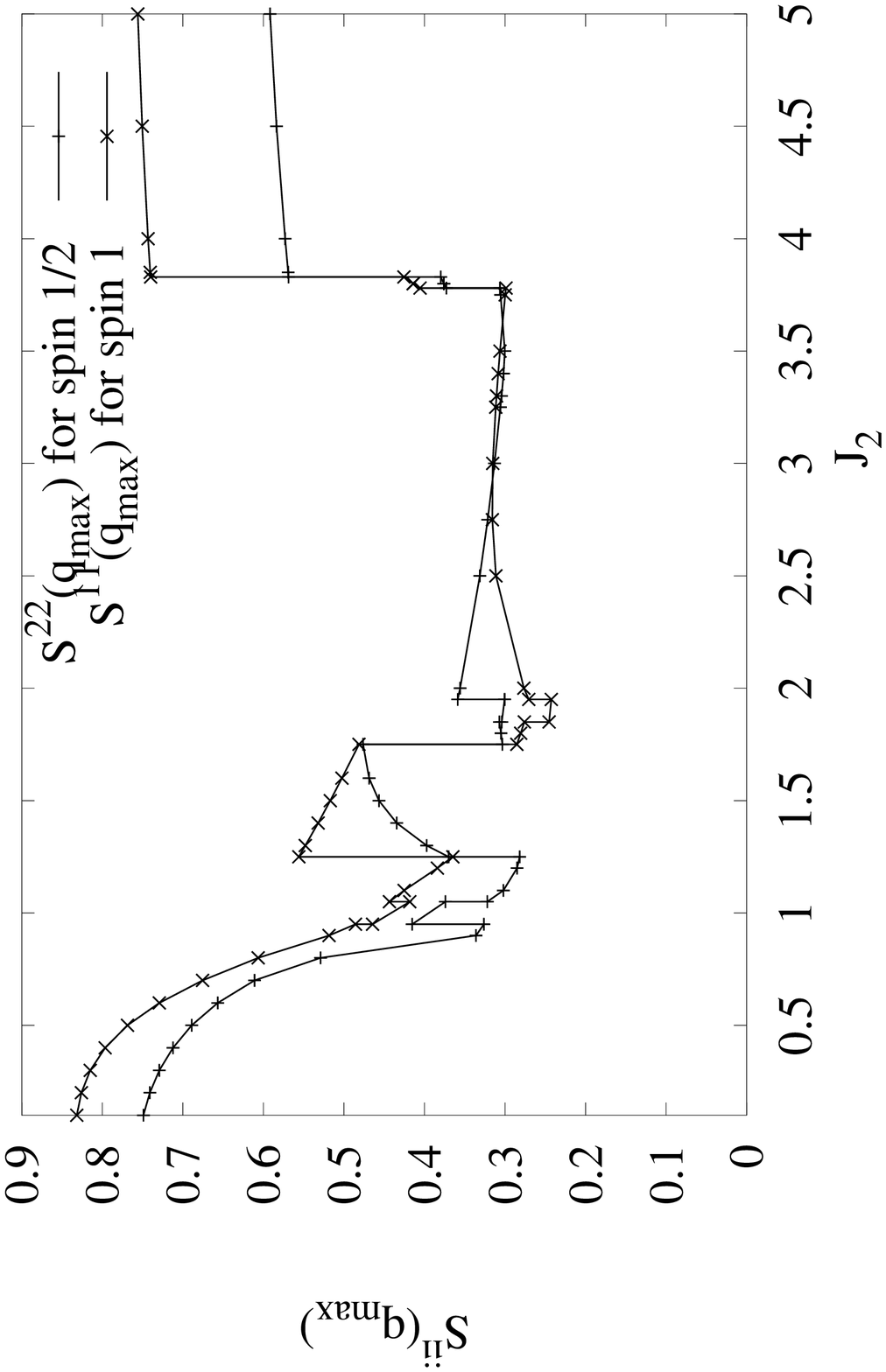,width=13cm}
\end{center}
\end{figure}
\centerline{Fig. 11}

\end{document}